\documentclass[%
 reprint,
superscriptaddress,
amsmath,amssymb,
 aps,
prb,
]{revtex4-2}

\usepackage[utf8]{inputenc}
\usepackage[caption=false]{subfig}
\usepackage{graphicx}
\usepackage{amsmath, amssymb}
\usepackage{bm}
\usepackage{xcolor}

\usepackage[english]{babel}

\usepackage{hyperref}

\usepackage{siunitx}

\DeclareMathOperator{\Tr}{Tr}
\DeclareMathOperator{\sinc}{sinc}

\begin{document}

\title{Magnetoelastic conversion in integrated YIG nanostructures}
\author{A.~V.~Bondarenko}
\affiliation{Kavli Institute of Nanoscience, Delft University of Technology, Delft, 2628CJ, the Netherlands}
\affiliation{Institute of Magnetism of NAS and MOS of Ukraine, Kyiv, 03142, Ukraine}
\author{F.~Engelhardt}
\affiliation{Institute for Theoretical Solid State Physics, RWTH Aachen University, Aachen, 52074, Germany}
\affiliation{Max Planck Institute for the Science of Light, Staudtstr. 2, PLZ Erlangen 91058, Germany}
\author{M.~Kounalakis}
\affiliation{Institute for Theoretical Solid State Physics, RWTH Aachen University, Aachen, 52074, Germany}
\affiliation{Luxembourg Institute of Science \& Technology (LIST), 4362, Esch-sur-Alzette, Luxembourg}
\author{T.~Valet}
\affiliation{Universit\'{e} Grenoble Alpes, CEA, CNRS, Grenoble INP, Spintec, Grenoble, France}
\author{O.~Klein}
\affiliation{Universit\'{e} Grenoble Alpes, CEA, CNRS, Grenoble INP, Spintec, Grenoble, France}
\author{G.~E.~W.~Bauer}
\affiliation{WPI-AIMR, Tohoku University, 2-1-1 Katahira, Sendai 980-8577, Japan}
\affiliation{Kavli Institute for Theoretical Sciences, University of the Chinese Academy of Sciences, Beijing, China}
\author{S.~{Viola Kusminskiy}}
\affiliation{Institute for Theoretical Solid State Physics, RWTH Aachen University, Aachen, 52074, Germany}
\affiliation{Max Planck Institute for the Science of Light, Staudtstr. 2, PLZ Erlangen 91058, Germany}
\author{Ya.~M.~Blanter}
\affiliation{Kavli Institute of Nanoscience, Delft University of Technology, Delft, 2628CJ, the Netherlands}

\begin{abstract}
Motivated by the recent proposal of two-step transduction from microwave to optical domain using magnetic and elastic intermediate stages \cite{Engelhardt}, we consider the coupling between resonant magnetic and elastic modes within a simple axially-symmetric nanodevice designed to host high-quality-factor acoustic modes: A suspended YIG ring structure supported by a central stem, fabricated from a continuous single-crystal film. We study the modes of the system with our custom finite element solvers. We identify the lowest order ``breathing'' mode of a magnetic vortex and the lowest order elastic breathing mode as having the largest mode overlap. For this pair of modes, the external out-of-plane magnetic bias field is critical for bringing them into resonance; however, we show that at the same time it also affects the strength of the coupling. To counteract this, we optimize the radius of the ring at fixed thickness. For the \qty{100}{\nano\meter}-thick film the \emph{resonant} coupling is maximized at $g/2\pi = \qty{8}{\mega\hertz}$ at $R\approx\qty{1.7}{\micro\meter}$, indicating that the overlap integral approaches the idealized limit assumed in previous order-of-magnitude estimates. Our results pave the way for the design of tunable frequency-conversion devices based on magnetoelastics.
\end{abstract}
\maketitle

\section{Introduction}
Efficient conversion of electromagnetic signals between microwave and optical domains remains one of the central challenges in transduction physics. Coherent transduction specifically yields significant improvements even in the classical regime, where coherent fiber interfaces are now pushing bandwidths into the \qty{1}{\tera\bit} range. In the quantum regime, coherent conversion, namely the bidirectional transfer of a quantum state between a qubit and an optical photon~\cite{Lambert,Lauk}, enables qualitatively new applications by combining long-range communication at infrared or optical frequencies with microwave (MW) frequency operated quantum computation platforms, such as superconducting qubits, neutral atoms, and trapped ions.



For coherent microwave-to-optics transduction, various (classical) transduction schemes were proposed in the literature. Direct electromagnetic to electromagnetic (electro-optical) conversion is possible using non-linear crystals with a conversion efficiency of around 0.1~\cite{Sahu}. Other proposals include an intermediate state. To justify introducing such a state, a sufficient gain in efficiency for both interconversions to an intermediate state is required. For the intermediate one-step conversion several schemes have been proposed, including optomechanical conversion, which uses a mechanical resonator as an intermediary, see {\em e.g.} Refs.~\onlinecite{Bochmann,Bagci,Regal2014,Forsch,Stockill2022}, and magnonic conversion~\cite{Hisatomi,Zhu2020}, which works with magnons --- the quanta of spin waves, the elementary excitations of a magnetic structure. For the time being, the field remains at the proof-of-concept stage. While optomechanical conversion has achieved high efficiency, around 0.5~\cite{Regal2018}, it suffers from the low bandwidth $\approx\qty{10}{\kilo\hertz}$ when tuned for optimal peak efficiency, whereas magnon-assisted conversion suffers from the low efficiency due to weak coupling between magnons and optical photons~\cite{Rameshti}.

For multistage approaches in particular, in addition to the coupling strength, one needs to make sure that all systems taking part in the conversion are also sufficiently long-lived. This is typically quantified by the cooperativity, which for a pair of coupled systems A and B is defined as $C = g^2_{AB}/4\gamma_A \gamma_B$. Here, $g_{AB}$ is the strength of the coupling between the systems, and $\gamma_{A,B}$ are the linewidths resulting from the losses. The cooperativity directly influences important measures of quantum performance, like the fidelity, and generally high cooperativities are desirable to maximize the quantum performance.

Recently,~\cite{Engelhardt} theoretically proposed a two-step conversion scheme which ideally would result in both higher efficiency and higher bandwidth. The scheme combines the advantages of both optomechanical and magnonic transduction --- optimized coupling of mechanical resonators to optical light and strong coupling of magnons to MW cavity photons. In the scheme, a MW photon couples to a magnon, the magnon couples resonantly to a phonon --- an elementary excitation of the elastic mode of the same magnetic microstructure that hosts the magnon, --- and lastly the phonon couples to the optical photon. The basics of all transduction steps are understood: Magnon-photon coupling is studied in cavity magnonics (see e.g. Ref.~\onlinecite{Rameshti}) and phonon-photon interaction in optomechanics~\cite{Aspelmeyer_2014}. The magnon-phonon coupling, due to the magnetoelastic interaction, has also been studied recently, and it was demonstrated to be strong and to lead to coherent state transfer between phonons and magnons~\cite{An2020,Schlitz,Hatanaka}. The currently preferred material for magnonics is Yttrium Iron Garnet (YIG), which is a ferrimagnetic insulator with high magnetic quality, transparent in the infrared range, low mechanical damping and high sound velocity. By co-optimizing the individual conversion stages, Ref.~\onlinecite{Engelhardt} showed that high efficiency over a broad frequency band can be achieved without requiring all cooperativities to be simultaneously matched; rather, control over only a subset of them suffices for the overall conversion efficiency to be maximized. Furthermore, for YIG, Ref.~\onlinecite{Engelhardt} produced simple estimates assuming that microwave photons, magnons, and phonons are all tuned to a resonance, in particular it showed that the magnon-phonon coupling strength can be in the MHz range, and the total efficiency can reach $0.9$, with a bandwidth of $0.5$ MHz.

\begin{figure}
    \centering\includegraphics[width=\columnwidth]{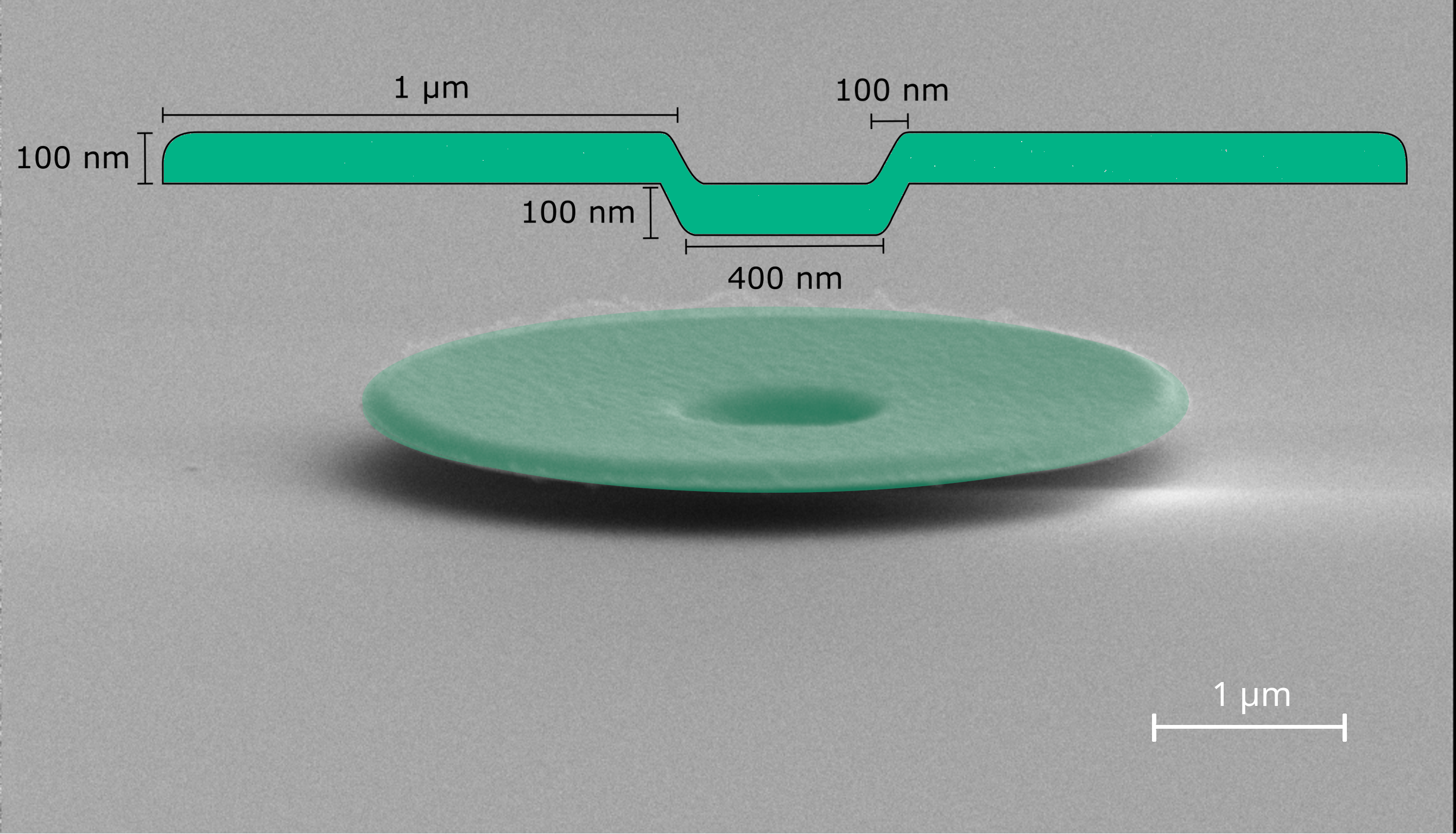}
    \caption{\label{fig:structure} Scanning Electron Microscope (SEM) image of a suspended magnetic YIG (yttrium iron garnet) ring mounted on a supporting disk stem. The structure is a single crystal oriented in the (111) direction. Image courtesy of the G. Schmidt group from the University of Halle~\cite{Heyroth2019}. The top inset presents a schematic cross-sectional view illustrating the nanostructure’s profile used in the simulations by finite element methods. 
    }
\end{figure}

Further research is needed to outline realistic device geometries at which these efficiency and bandwidth can be realized. Specifically for magnon-phonon coupling, which is the focus of this manuscript, we need to identify the optimal geometry and the relevant magnon and phonon modes. To this end, we show how to optimize the resonant magnon-phonon coupling in a microdisk magnetic structure hosting a magnetic vortex, which is a promising setup to achieve the ultimate goal of high transduction efficiency. On the one hand, microdisk resonators are a standard structure in cavity optomechanics~\cite{matsko_optomechanics_2009, savchenkov_surface_2011, ding_wavelength-sized_2011}. They can host optical whispering gallery modes (WGM) that interact with mechanical deformations of the boundary modulating the optical pathlength, resulting also in good values for the optomechanical coupling strength. On the other hand, in the past years, progress has been made in the fabrication of micro-sized disk resonators made of magnetic materials, in particular YIG~\cite{losby_torque-mixing_2015, zhu_patterned_2017, collet_generation_2016, srivastava_identification_2023, hamadeh_core_2024, seeger_experimental_2024}, furthermore, the disks can be mechanically suspended by a narrow central stem (see Fig.~\ref{fig:structure}) to reduce phonon linewidths in these structures~\cite{Schmidt1,Schmidt2,nguyen_ultrahigh_2013,nguyen_improved_2015}. Magnets of small size are known to have a non-trivial ground state of the magnetization, a vortex state~\cite{shinjo_magnetic_2000, guslienko_magnetic_2008}. This motivated the first proposal to couple magnetic excitations in these geometries to optical WGMs via the Faraday effect~\cite{Graf2018}. Whereas the possibility of tuning and enhancing the optomagnonic coupling in these structures was shown, the estimated maximum cooperativity is still too low for a single-stage transduction. The axial symmetry of these structures makes them also a reasonable choice for optimizing the magnetoelastic coupling between suitably chosen modes, as we show below, providing a platform for a two-step transduction scheme as proposed in Ref.~\cite{Engelhardt}.

With these considerations in mind, we identify optimal magnetic and elastic modes in the proposed integrated magnetic–mechanical nanoresonator. Then we show that the magnon-phonon coupling strength, with a suitable choice of parameters, indeed can be brought to the MHz range, in accordance with the estimates of Ref.~\cite{Engelhardt}. Such coupling should be sufficient to match the cooperativities of microwave -- magnon and optomechanical couplings and thus to facilitate the two-stage microwave-optical transduction.

This manuscript is organized as follows: In Sec. \ref{sec:model} we introduce the model, giving the Hamiltonian and the general expression for the magnetoelastic coupling. In Sec. \ref{sec:elastic} and Sec. \ref{sec:magnetic} we present the results for elastic and magnetic modes for the ring geometry, respectively. In Sec. \ref{sec:coupling} we study the magnetoelastic coupling. Sec. \ref{sec:conclusions} contains the conclusions and outlook.

\section{Model} \label{sec:model}

We now formulate a coupled magnetic–elastic description of the suspended YIG ring introduced above. The system, possessing both magnetic and elastic degrees of freedom, is described by the following Hamiltonian density, 
\begin{equation}
    \mathcal{H}=\mathcal{H}_\text{mag}[\bm{m}]+\mathcal{H}_\text{el}(\bm{u}) + \mathcal{H}_\text{int}(\bm{m},\bm{u}) \ .
\end{equation}
Here the displacement field $\bm{u}(\bm{r},t)$ is defined via $\bm{r}'(\bm{r},t)=\bm{r}+\bm{u}(\bm{r},t)$, describing the deformation of the infinitesimal volume element originally located at $\bm{r}$, and the local magnetization field is given by $M_s\bm{m}(\mathbf{r}, t)$, with $M_s$ being the saturation magnetization and $\bm{m}(\mathbf{r}, t)$ a unit vector. The contributions $\mathcal{H}_\text{mag}$ and $\mathcal{H}_\text{el}$ describe the magnetic and elastic subsystems, respectively, while $\mathcal{H}_\text{int}$ is the density of their interaction energy. In the following, we treat the magnetoelastic interaction as a perturbation. Having in mind that YIG is typically grown on a Gadolinium Gallium Garnet (GGG) substrate, we consider the YIG crystal oriented such that the (111) crystallographic direction is perpendicular to the plane of the ring.

We disregard non-linear elastic effects and consider the elastic part of the Hamiltonian to be quadratic. We also linearize the magnetic part, see Appendices~\ref{ap:magnon_norm} and ~\ref{ap:phonon_norm}. In a confined geometry like the one we consider, both magnetic and elastic spectra are discrete. The general solutions for the unperturbed magnetic and elastic Hamiltonians are linear combinations of the respective eigenmodes, 
\begin{align}
    \bm{m}&=\bm{m}_0 + {\rm Re}\left[\sum_j m_j \bm{m}^{(d)}_j e^{-i \omega_j t}\right],\label{eq:mag_anzats}\\
    \bm{u}&={\rm Re} \left[\sum_k b_k \bm{u}_k e^{-i \omega_k t}\right],
    \label{eq:el_anzats}
\end{align}
where $j$ and $k$ are the indices describing magnetic and elastic modes, respectively, $\bm{m}_0$ is the static magnetization, $\bm{m}^{(d)}_j(\bm r)$ and $\bm{u}_k(\bm r)$ are the complex mode profiles satisfying the linearized equations of motion, $\omega_{j}, \omega_{k}$ are the mode frequencies, and $m_j,\;b_k$ are the complex-valued amplitudes of the modes. In the linear limit, we write the Hamiltonian integrated over the whole volume as
\begin{equation}
    H=\int \mathcal{H} dV = \sum_j \alpha_j m^\ast_j m_j + \sum_k \beta_k b^\ast_k b_k + \int \mathcal{H}_\text{int} dV,
\end{equation}
where we have omitted the constant term, and the coeficients $\alpha_j$ and $\beta_k$ are dependent on the normalization of the amplitude, derived from Eqs.~\eqref{eq:mag_norm} and~\eqref{eq:phon_norm},
\begin{align}
\alpha_j&= \omega_j \frac{M_s}{\gamma_0}\int dV \bm{m}_0 \left[{\rm Im}(\bm{m}^{(d)}_j)\times{\rm Re}(\bm{m}^{(d)}_j)\right],\\
\beta_k&=\rho\omega_k^2 \int dV \bm{u}_k^\ast \bm{u}_k ,
\end{align}
where $\rho$ is the YIG mass density and $\gamma_0$ the effective gyromagnetic ratio.
The complex amplitudes $m_j$ and $b_k$ which appear in the Hamiltonian form canonical variable pairs together with their respective conjugates, and the equation of motion for the reduced system can now be treated as a system of ordinary differential equations for the complex amplitudes. 

The magnetoelastic energy density $\mathcal{H}_\text{int}$ is, to leading order, linear in the strain tensor and bilinear in the magnetization components. In particular, it has the simplest form for a cubic crystal in the (100) orientation~\cite{PhysRev.110.836},
\begin{align}
\mathcal{H}_\text{int}&=B_{ijkl}\varepsilon_{ij}m_k m_l =\notag\\
    &=B_1 \left(m_x^2 \varepsilon_{xx} + m_y^2 \varepsilon_{yy} + m_z^2 \varepsilon_{zz}\right)+\notag\\
    &+2B_2 \left(m_x m_y \varepsilon_{xy} + m_x m_z \varepsilon_{xz} + m_y m_z \varepsilon_{yz}\right) ,
    \label{eq:100ms_energy}
\end{align}
where $B_{1,2}$ are the two independent elements of the magnetostrictive tensor of a material with cubic symmetry $B_{11}$ and $B_{44}$~\cite{PhysRev.110.836, Vanderveken2021} in Voigt notation, respectively, and $\varepsilon_{ij}$ are the components of the strain tensor.

In this Article, we need the expression in the (111) orientation, for which the coordinate frame must be rotated as described in Appendix~\ref{app:tens_der}. We write here the resulting expression in cylindrical coordinates,
\begin{align} \label{magnetostriction_interaction}
\mathcal{H}_\text{int}=&\frac{B_1+2B_2}{3}\left(m_\phi^2 \frac{u_r}{r}+m_r^2 \frac{\partial u_r}{\partial r}\right)+\notag\\
    &+\frac{B_1-B_2}{6}m_z^2\left(\frac{u_r}{r}+\frac{\partial u_r}{\partial r}\right)+ B_2 m_z^2 \frac{\partial u_z}{\partial z} +\notag\\
    &+\frac{2B_1+B_2}{3}m_r m_z \left(\frac{\partial u_r}{\partial z}+\frac{\partial u_z}{\partial r}\right)+F(\mathbf{u}),
\end{align}
where the magnetization-independent term $F(\mathbf{u})$ results from imposing $\bm{m}^2=1$. 

Linearizing the interaction around the  equilibrium magnetization $\bm{m}_0$, we obtain a linear coupling $\mathcal{H}_\text{int} \sim m_j^\ast b_k +\text{c.c.}$ Renormalizing the  amplitudes by the replacements $m_j\rightarrow \hbar \omega_j m_j/\sqrt{\alpha}_j,\;  b_k\rightarrow \hbar \omega_k b_k/\sqrt{\beta_k}$, we bring our Hamiltonian to the conventional form, 
\begin{eqnarray}
    H & = & \sum_j \hbar\omega_j m^\dagger_j m_j + \hbar\sum_k \omega_k b^\dagger_k b_k \nonumber \\
    & + & \hbar\sum_{j,k} \left( g_{jk}m_j^\dagger b_k + g_{jk}^\ast m_j b_k^\dagger\right),
\end{eqnarray}
where the coupling constant $g_{jk}$ is computed using the now normalized mode distributions,
\begin{equation}
    \hbar g_{jk} = \frac{1}{\sqrt{\alpha_j\beta_k}}\int dV B_{\alpha \beta \gamma \delta} u_{k,\alpha\beta} m_{0\gamma}m^{\ast(d)}_{j,\delta},
    \label{eq:coupling_overlap}
\end{equation}
with summation implied over repeated spatial indices $\alpha,\beta,\gamma,\delta \in {x,y,z}$.

In the remainder of the manuscript, we describe the magnon modes, the elastic modes, and use Eq. (\ref{eq:coupling_overlap}) to calculate and characterize the coupling constant. Since the considered geometry is axially symmetric, we can separate the azimuthal angle $\varphi$ from the rest of the problem and write both magnetic and elastic modes in the form $f(r,z)e^{i m \varphi}$. This greatly reduces the complexity of the problem. In particular, both the Zeeman coupling of the magnetic structure to the external magnetic field, $-\mathbf{M}\cdot\mathbf{B}$, and the magnetoelastic overlap integral Eq.~\eqref{eq:coupling_overlap} explicitly enforce orthogonality with respect to the azimuthal index. --- only modes with the same azimuthal number can be coupled linearly. Since the relevant driving fields are spatially uniform on the scale of the structure, only azimuthally symmetric ($m=0$) modes are efficiently excited. We therefore restrict our analysis to these modes.

For the confined geometry, we need to introduce two more discrete indices --- a radial number $n$ and the axial number $n_z$. Due to the large aspect ratio of the structure, the frequency spacing between successive axial modes is significantly larger than that between radial modes, hence we also limit ourselves to study the $n_z=0$ modes.

\section{Elastic modes} \label{sec:elastic}
We calculate the elastic modes based on the standard theory of elasticity. Our starting point is the equation of motion for the components of the displacement field $\bm{u}(\bm{r},t)$,
\begin{equation}
    \rho \ddot{u}_{\alpha} = \partial_\beta C_{\alpha\beta\gamma\delta}\varepsilon_{\gamma\delta},
\end{equation}
$\rho$ is the mass density, and the stiffness tensor $\hat{C}$ is taken to be that of an isotropic material. Elastic anisotropy can, in principle, be included perturbatively, as it is small for YIG (the anisotropy factor $2C_{44}/(C_{11}-C_{12})=0.947$~\cite{Clark}).

\begin{figure}[t]
\centering\includegraphics[width=\columnwidth]{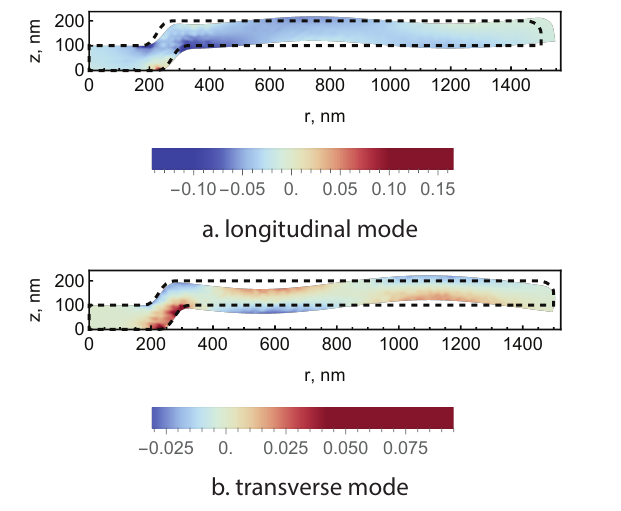}

\caption{EIGENELAAxi FEM simulation of the $n_r=0$ elastic breathing mode (a) and the transverse mode closest to it in frequency (b) in the structure from Fig.~\ref{fig:structure}. The structural central attachment stem causes the longitudinal oscillation to have a significant out-of-plane component compared to the idealized planar structure. The cross-section of the (exaggerated) deformation is compared to the non-deformed profile (outlined with the dashed line). The color represents a hydrostatic strain $\Tr \varepsilon\equiv\nabla \mathbf{u}$ due to the local compression and expansion. Globally the structure is depicted in the extension phase of the breathing oscillation. For a fundamental purely longitudinal mode, we would only have tensile stress in this phase, but we also see small pockets of compressive stress because the longitudinal mode is coupled to the transverse mode.}
\label{fig:FEM_solver_elastic}
\end{figure}

\begin{figure}
    \centering\includegraphics[width=\columnwidth]{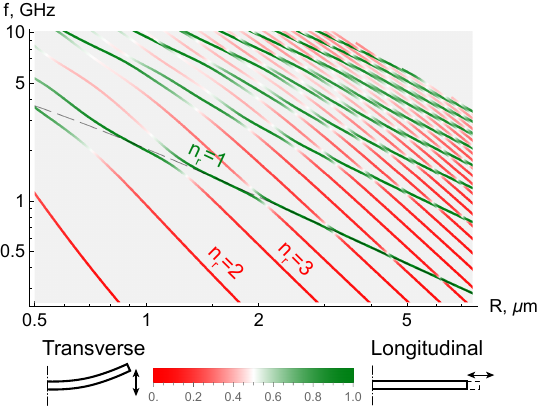}
    \caption{\label{fig:mechanics-modes} Frequencies of the mechanical modes of the structure shown in Fig.~\ref{fig:structure}. Torsion $u_\varphi\ne0$ modes have been omitted. The color gradient in green and red denotes the ratio between the average in- and out-of-plane displacements $\sqrt{\int u_r^\ast u_r dV} / \sqrt{\int u_z^\ast u_z dV}$; the ratio is used to describe the modes in terms of familiar longitudinal and transverse modes observed in planar geometries. The dashed line shows the analytical behavior of the fundamental $n_r=1$ longitudinal mode for a disk with a \qty{300}{\nano\meter} diameter hole in the middle for comparison.}
\end{figure}
For a symmetric system with a high aspect ratio, the elastic equations typically produce distinct longitudinal (deformation parallel $\bm{u}\parallel\bm{e}_r$) and transverse ($\bm{u}\perp\bm{e}_r$) modes~\cite{Landau1986}. However, if the geometry becomes more complex, like in our case, the two cannot be clearly separated.

In Fig.~\ref{fig:FEM_solver_elastic} we show the displacement map of the ring on a stem produced by the custom solver, EIGENELAAxi, developed by CEA and based on the finite element method (FEM). For the simulations we use an isotropic model with the Young's modulus $E=\qty{198}{\giga\pascal}$, Poisson's ratio $\nu=0.3$, and mass density $\rho=\qty{5170}{\kilogram\per\meter^3}$. Conversion to the elastic tensor $C$ is done with a procedure described in Ref.~\cite{Mavko_Mukerji_Dvorkin_2009}. The plot shows that longitudinal and transverse oscillations are mixed, though the contribution of transverse ones is relatively weak, since the deformation in the radial direction is significantly larger than in the axial one.

Fig.~\ref{fig:mechanics-modes} shows the frequencies of elastic modes produced by our custom solver as a function of the radius of the structure. For reference, we also show an analytical solution for a ring without a stem (Appendix D), which approximately behaves as $\omega \propto 1/R$ with a dashed line. In the presence of the stem, we no longer have purely longitudinal and transverse modes; instead the modes have both in-plane and out-of-plane motion at the same time. However, it is possible to minimize one type of motion by choosing radii where the modes in a stemless disk are far from crossing each other, i.e., away from avoided crossings between longitudinal- and transverse-like branches. The lowest frequency mode is the radial breathing mode we are most interested in. As expected from geometric scaling, the breathing-mode frequency decreases with increasing outer radius and remains in the low-GHz range for all radii considered.

\section{Magnetic modes} \label{sec:magnetic}
\begin{figure}[t]
\subfloat{\includegraphics[width=0.45\textwidth]{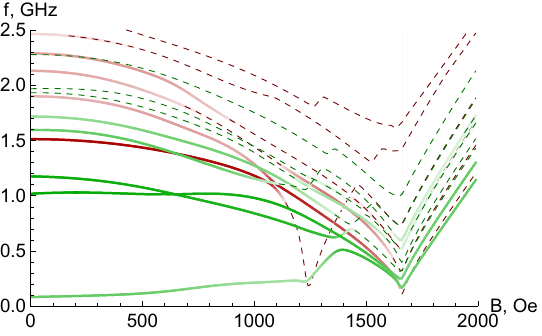}}
\caption{\label{fig:magnetic_modes}Simulated evolution of the ferromagnetic resonance (FMR) modes as a function of the external magnetic field \( B \) applied perpendicularly to the plane of the disk, computed using a custom-developed axisymmetric finite element eigensolver. The device corresponds to the suspended YIG disk shown in Fig.~\ref{fig:structure}. Modes obtained via FEM are classified based on their excitation symmetry: modes predominantly coupled to an in-plane AC field (azimuthal symmetry \( m = 1 \)) are shown in green, while those coupled to an out-of-plane AC field (\( m = 0 \)) are shown in red. 
The line opacity shows the magnitude of the total dynamic dipolar moment of the mode $\left|\int dV\, \bm{m}^{(d)}\right|$, which is directly proportional to the coupling strength of a given mode to a microwave cavity; modes that lose their coupling are continued with dashed lines. 
As the external field increases, the magnetic texture continuously deforms from the vortex state through the cone state and into the saturated uniform state. The common mode softening point indicates the transition field \( B_s \approx \qty{1650}{Oe}\), at which the magnetization switches from a cone state to a uniform state. Notably, for certain disk radii, such as the one presented here, an additional transition is observed at a slightly lower field than \( B_s \), manifested by the softening of only a limited number of modes. This transition corresponds to an independent saturation of the stem and affects modes that are effectively pinned to it.
}

\end{figure}

In this Section, we calculate the magnetic modes with our custom SLLGAxi (developed by CEA) finite element solver. We adopt the reference values for magnetic constants of YIG $M_s=\qty{139}{\kilo\ampere\per\meter}$ and $A_\text{ex}=\qty{3.7}{\pico\joule\per\meter}$. Magnetocrystalline anisotropy is omitted, since the present axially-symmetric solver does not allow to it to be included directly; its contribution can instead be incorporated perturbatively in post-processing.

First, we discuss the dependence on the bias magnetic field. The ground-state magnetization saturates out of plane for external fields slightly smaller than the thin-film shape anisotropy field $B_s \lesssim  4\pi M_s$ (the difference between the two becomes significant only when the overhanging ring is much smaller than the stem), and all the modes exhibit the softening near the saturation point, significantly reducing their frequencies at the critical field $B_s$. For $B < B_s$ the equilibrium configuration is a magnetic vortex and the related cone state~\cite{PhysRevB.65.134434}, see below. For $B > B_s$ we have a homogeneously magnetized sample with the well-studied family of so-called drum modes~\cite{Kakazei,Dobtrovolskiy}, which scale linearly with the external field. 

In this Article, we are interested in the vortex modes below $B_s$. The vortex state is characterized by the spins orienting themselves in circles such that on the circular edge they are tangential to the surface~\cite{doi:10.1126/science.1075302}, the dipolar interactions are strengthened, and the mode frequencies are high even without an external bias field. A distinction has to be made depending on whether the modes involve the motion of the vortex core~\cite{PhysRevB.58.8464,PhysRevLett.94.027205}. When a magnetic bias field is applied out of plane the deformed vortex state is also known as the cone state~\cite{PhysRevB.65.134434}, characterized by moments being pulled out of plane.

The modes which move the vortex core are referred to as gyrotropic modes~\cite{10.1063/1.1652420,D4NH00145A}. They serve as the fingerprint of the vortex state which facilitates its experimental observation. However, they are also highly localized and interact with phonons much more weakly than non-gyrotropic modes because of the limited overlap with the phonon modes. The gyrotropic modes can be coupled directly to optical WGMs but with predicted cooperativities lower than 1~\cite{Graf2018}.

In Fig.~\ref{fig:magnetic_modes} we see both the in-plane, shown in green, and out-of-plane, in red, excitations. The in-plane response is strongest for the non-trivial axial orders $n_z\ne0$ of gyrotropic modes, see Ref.~\cite{D4NH00145A}. However, much more interesting to our purpose is the strongest out-of-plane excited mode shown in the darkest red, which has all the periphery spins in motion around the circular static magnetization of the vortex. The precession of the spins near the vortex core also causes the core size to change periodically, so we will refer to it as a vortex breathing mode (VBM) in analogy to the one in skyrmions~\cite{PhysRevB.90.064410}. All the other observable out-of-plane excited modes are radial orders $n_r$ of this fundamental vortex breathing mode.

\begin{figure}[t]
    \includegraphics[width=\columnwidth]{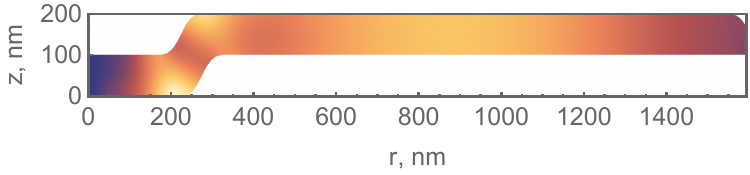}
    \centering\includegraphics[width=0.5\columnwidth]{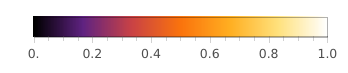}
    \caption{\label{fig:magnon_distrib}Calculated spatial distribution of the vortex breathing mode intensity $\left[\mathbf{m}_0\times\mathbf{m}^{(d)\ast}\right]\mathbf{m}^{(d)}$ using the SLLGAxi FEM solver with external field $B=0.1$~T. The stem defines three regions, viz. the suspended ring, the joint, and the circular base of the structure.}
\label{fig:mumax_FEM}
\end{figure}

From the magnetic mode intensity distribution on Fig.~\ref{fig:magnon_distrib} we can see that, unlike the elastic modes, the magnetic mode we are interested in is localized primarily in the suspended ring and secondarily in the flat area of the stem. Similarly to the curvature producing localized demagnetizing fields for 1D magnets~\cite{makarov2022curvilinear}, the inhomogeneities of the demagnitizing field supress mode in the curved connecting area. Thus, the stem, which was added to reduce mechanical losses, has little effect on the magnetic properties of the ring.

\section{Magnon-phonon coupling} \label{sec:coupling}

\subsection{Resonant matching}
\begin{figure}[t]
\centering\includegraphics[width=0.9\columnwidth]{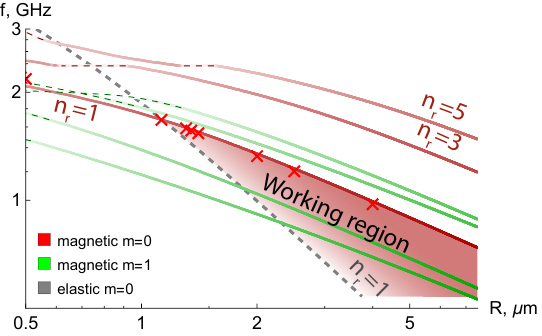}
\caption{The frequencies of the magnetic modes with the best coupling to the cavity compared to the fundamental elastomechanical mode of the system versus the radius of the ring. As previously, colour denotes in-plane excited $m=1$ modes in green and out-of-plane $m=0$ modes in red. The elastic mode frequencies are an analytical fit for the fundamental in-plane modes from Fig.~\ref{fig:FEM_solver_elastic} using the model of a disk with a hole at the center ($t=\qty{100}{\nano\meter},\;r_0=\qty{200}{\nano\meter}$), and the magnetic modes were calculated for the realistic structure in the custom FEM solver. The crosses are the frequencies obtained from Mumax3 simulations in the same realistic geometry, see Appendix \ref{ap:mumax}. The crossing point is found at $R\approx \qty{1.3}{\micro\meter}$.}
\label{fig:freq_overlap}
\end{figure}

Before evaluating the magnetoelastic coupling, we discuss the conditions under which the magnetic and elastic modes can be tuned into resonance, where the interaction is expected to be strongest. The modes scale differently with the ring radius, and since both lie in the low-GHz range, resonance can be achieved by adjusting the radius and by applying an external magnetic bias field. In Fig.~\ref{fig:freq_overlap} we show the frequencies of a few lowest elastic and magnetic modes in zero magnetic field as a function of the radius of the ring. The crossing occurs at practical micron-scale radii. In a finite bias field, the elastic modes remain the same while the magnetic modes go down in frequency. Therefore in practice, having a radius larger that that of the crossing point (at zero field) is desirable because one can tune the modes into the resonance to compensate for possible fabrication variations.

The lowest radius at which we can find a crossing magneto-elastic pair occurs for the fundamental ($n_r=0$) magnetic and elastic modes, higher order modes cross at much larger radii.

\subsection{Optimizing the mode overlap}

We now calculate the overlap between magnetic and elastic modes and discuss how it can be optimized. Our starting point is Eq.~\eqref{magnetostriction_interaction} for the magnetoelastic part of the Hamiltonian in the (111) YIG orientation together with Eq.~\eqref{eq:coupling_overlap} for the coupling strength.

Before presenting the numerical results in the next Subsection, we provide a qualitative discussion that offers intuitive insight into the coupling mechanism and demonstrates that an out-of-plane bias magnetic field is required to obtain a significant coupling strength. 

We first simplify the problem by ignoring the contribution of the vortex core region, which is secondary for our mode selection, and only consider the periphery of the vortex. For completeness, we provide estimates of the vortex core contribution to the coupling in the disk geometry in Appendix \ref{appendix:vortex_core} and show that it is negligible compared to the periphery contribution considered below.

To make the estimate, we model the equilibrium magnetization of the vortex as 
\begin{equation}
    \bm{m}_0 = \sin \alpha \cos \theta \bm{e}_r + \cos \alpha \cos \theta \bm{e}_\phi + \sin \theta \bm{e}_z,
\end{equation}
where $\bm{e}_r,\bm{e}_\phi,\bm{e}_z$ are the local cylindrical basis vectors at position $\bm{r}$. Furthermore, $\theta$ is the angle of the spins with respect to  the plane (to describe the cone-state, see Ref.~\cite{PhysRevB.65.134434}), and $\alpha$ describes the helicity of a vortex~\cite{Verba2020}.

A pair of vectors perpendicular to $\bm{m}_0$  is then needed to form a basis for modeling the linear oscillations. We choose $\bm{m}_1=\begin{pmatrix}
    \cos \alpha & -\sin \alpha  & 0
\end{pmatrix},$ and $\left[\bm{m}_1 \times \bm{m}_0\right]$ perpendicular to both $\bm{m}_0$ and $\bm{m}_1$. This yields the following model for the precession,
\begin{equation}
    \bm{m}^{(d)}=m_\text{amp}\left(\sqrt{1-\epsilon^2} \left[\bm{m}_1 \times \bm{m}_0\right] + i \bm{m}_1\right),
\end{equation}
where $m_\text{amp}$ is the spin-precession amplitude, and $\epsilon$ describes the eccentricity of the ellipse traced by the precessing magnetization($0 \le \epsilon < 1$). The ellipticity originates from thin-film anisotropy; from numerical simulations we estimate $\epsilon \approx 0.6$. For simplicity, we further assume that the parameters $\alpha$, $\theta$, and $\epsilon$ are approximately constant within the suspended portion of the disk, in reasonable agreement with the numerical results.

We can produce analytical expressions for the magnetoelastic energy density $\mathcal{H}_\text{int}$ in two limiting cases,
\begin{eqnarray}
    \theta=0:\;& \mathcal{H}_\text{int}  = & \frac{B_1+2B_2}{3} \sin 2\alpha \left(\frac{\partial u_r}{\partial r}-\frac{u_r}{r}\right) m_\text{amp.},
    \nonumber \\
    \alpha=0:\;& \mathcal{H}_\text{int}  = & \frac{B_1+5B_2}{6} \sqrt{1-\epsilon^2} \sin 2\theta \frac{u_r}{r} m_\text{amp.}, \label{eq:oop_enhance} 
\end{eqnarray}
Note that we use a purely longitudinal Ansatz ($u_z=0$) to derive Eqs. (\ref{eq:oop_enhance}) and thus effectively retain only the first two terms of Eq. (\ref{magnetostriction_interaction}). We justify such a treatment by checking the precise shear contribution from the numerical results discussed later; the shear contribution does not exceed 10\%.

The key conclusion of the model estimates~\eqref{eq:oop_enhance} is that if the magnetization is tangential to the surface, $\alpha=0$, and at the same time localized in the plane, $\theta=0$, the periphery contribution to the magnetoelastic coupling vanishes. To enable this contribution, the structure of the vortex must be perturbed, for example, by the external out-of-plane field. To estimate the value of the coupling in the presence of the field, we resort to the numerical treatment.

It is also worth noting that both in numerical results and in the analytical expressions Eqs.~(\ref{eq:oop_enhance}) the contribution of the constant $B_2$ to the coupling is significantly larger than that of $B_1$. This is evident from the ratios in the prefactors of Eqs.~(\ref{eq:oop_enhance}). It is important since the value of the $B_2$ increases several times at cryogenic temperatures while $B_1$ goes down, see Refs.~\cite{Clark1963,Platzker}, which might be relevant for future studies of the system in the quantum regime. In the following, when we produce numerical results, both contributions are taken into account, and the room temperature values are used $B_1=\qty{3.48e5}{\joule\per\meter^3},$ and $B_2=\qty{6.96e5}{\joule\per\meter^3}$.

\subsection{Numerical results}

\begin{figure}
\centering\includegraphics[width=\columnwidth]{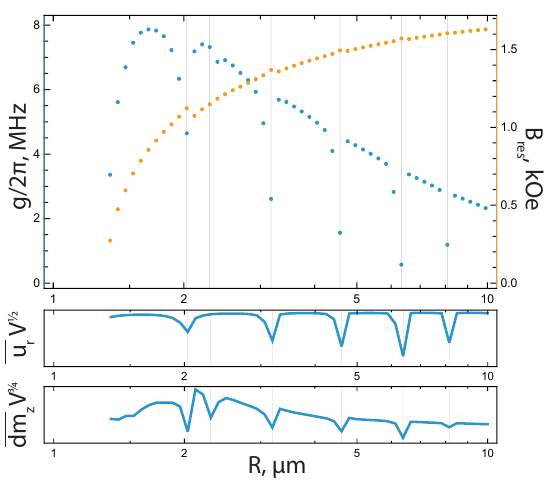}
\caption{\label{fig:coupl-v-bias} The coupling strength $g/2\pi$ and the tuning field $B_\text{res.}$ defined by $\;f_\text{mag.}(B_\text{res.})=f_\text{el.}$ versus the radius of the nanostructure. One can observe that there is an optimal radius at which the coupling is maximized since the tuning field approaches the field that maximizes the coupling, Eq.~\eqref{eq:oop_enhance}. The overall smooth trend is interrupted by a series of dips in the coupling strength associated with crossings of transverse elastic modes, seen on Fig.~\ref{fig:mechanics-modes}, and with another magnon mode, Fig.~\ref{fig:magnetic_modes}. 
These crossings manifest as dips in the average radial displacement $\overline{u_r}$ and in the average axial magnetic moment $\overline{dm_z}$.}
\end{figure}

We plot the dependance of the magnetoelastic coupling on the radius in Fig.~\ref{fig:coupl-v-bias} for the previously described structure. At every radius we find the field which tunes the magnetic mode frequency in resonance with the elastic mode, therefore only radii exceeding the previously identified critical value are shown.

As predicted by our analytical derivations, both the pure vortex state and the uniform state result in a weak coupling. There exists, however, a parameter window of ring radii at which the external field required to tune the system {\em also} optimizes the cone-state parameters for magnetoelastic interaction. For details on how the interaction strength varies with the external field, see Appendix~\ref{app:field_dep}. Experimental measurements of the magnetic ($\Gamma_m$) and elastic ($\Gamma_e$) damping rates of the suspended structure shown in Fig.~\ref{fig:structure} reveal that the corresponding linewidth broadening at 1~GHz is approximately 2~MHz. This opens up an opportunity to achieve \emph{coherent} hybridization between the magnetic and elastic modes when $g$ is greater than this value, which corresponds to the range of micron-size radius rings. 

The overall smooth trend is punctuated at a couple of radii where either (i) longitudinal elastic mode crosses the transverse mode, observable by a dip in the average radial displacement $u_r$, or (ii) the vortex breathing mode crosses a magnetic mode which produces a dip in $dm_z$ but not in $u_r$. Since the resonance condition is maintained by adjusting $B_\text{res}(R)$, the influence of these crossings appears in both observables, albeit with different relative strengths.

\section{Conclusions} \label{sec:conclusions}
In this work, we theoretically analyze the magnetoelastic properties of the suspended YIG ring platform experimentally realized in Ref.~\cite{Heyroth2019}, with the specific aim of assessing its suitability for coherent magnetoelastic conversion as proposed in Ref.~\cite{Engelhardt}. By resolving the full magnetic and elastic mode structure of this fabricated nanostructure, we determine the symmetry constraints, geometric conditions, and bias-field requirements necessary for achieving strong magnetoelastic coupling.

In our analysis we deliberately restrict the magnetic ground state to vortex and cone configurations. Although this choice was initially motivated by practical considerations, namely, their natural stability in the ring geometry and compatibility with axial symmetry, it turns out to be fundamentally advantageous. The non-uniform magnetization textures of vortex and cone states effectively relax the crystalline symmetry constraints that otherwise suppress magnetoelastic coupling, thereby enabling symmetry-allowed mode overlap and enhanced coupling strengths.

Using the custom FEM solvers developed by CEA, we compute the complete spectrum of magnetic and elastic eigenmodes of the suspended ring geometry, allowing for a systematic and exhaustive identification of candidate mode pairs. The magnetoelastic interaction imposes a strict selection rule on the azimuthal quantum number $m$, which severely constrains the set of symmetry-allowed couplings, whereas the selection rules associated with radial mode indices are comparatively less restrictive. 

Within this symmetry-filtered mode space, we identify a pair of fundamental breathing modes, magnetic and elastic, that can be tuned into resonance at experimentally realistic micrometer-scale radii. Higher-order radial overtones exhibit crossings only at substantially larger radii, rendering them less compatible with compact device geometries.

Although geometric freedom allows one to reach magnetoelastic resonance above a critical radius for a given mode pair, since the magnetic frequency can be tuned downward by the external bias field, resonance alone does not guarantee optimal coupling. We find that the symmetry of the magnetoelastic interaction plays a decisive role: the same out-of-plane field that brings the modes into resonance simultaneously reshapes the magnetization texture and thereby modifies the spatial overlap integral.

Consequently, device optimization requires selecting the ring radius such that the bias field satisfying the resonance condition also maximizes the symmetry-allowed mode overlap dictated by the crystalline structure of YIG. When this coordinated tuning of geometry and magnetic bias is achieved, the overlap approaches its theoretical maximum, corresponding to maximizing the effective interaction volume. For our chosen thickness of $t=\qty{100}{\nano\meter}$, this optimization yields a coupling strength of $g/2\pi=\qty{8}{\mega\hertz}$ at an optimal radius of $R\approx\qty{1.7}{\micro\meter}$. Strong magnetoelastic interaction in this platform therefore emerges not merely from frequency matching, but from symmetry-constrained co-optimization of geometry and bias field.



Ref.~\cite{Engelhardt} demonstrated that the transduction efficiency depends not on the absolute value of the coupling, but of the cooperativity, which needs to match in a certain way the cooperativities of the two other transduction steps --- microwave to magnon and elastic to optical conversion. The cooperativity also depends on the dissipation rates by magnons and phonons. Calculation of these dissipation rates is outside the scope of this manuscript, since it requires identification of relevant dissipation mechanisms. This, together with tackling the optomechanical coupling and optical dissipation for the YIG microstructure, are the next steps needed towards the full quantitative understanding of transduction in a realistic, axially symmetric device as proposed in this work.




\section{Acknowledgements}
This work was supported by the EU-project HORIZON-EIC-2021-PATHFINDER OPEN PALANTIRI-101046630; the French Grants ANR-21-CE24-0031 Harmony; and the PEPR SPIN - MAGISTRAL ANR-24-EXSP-0004; the Federal Ministry of Research, Technology and Space (BMFTR) project QECHQS (Grant No. 16KIS1590K). We thank G. Schmidt for providing the SEM image in Fig.\ref{fig:structure}.

\bibliography{YIG-coupling-references}

@article{nguyen_ultrahigh_2013,
	title = {Ultrahigh {Q}-frequency product for optomechanical disk resonators with a mechanical shield},
	volume = {103},
	issn = {0003-6951, 1077-3118},
	url = {https://pubs.aip.org/apl/article/103/24/241112/25885/Ultrahigh-Q-frequency-product-for-optomechanical},
	doi = {10.1063/1.4846515},
	language = {english},
	number = {24},
	urldate = {2025-03-13},
	journal = {Applied Physics Letters},
	author = {Nguyen, D. T. and Baker, C. and Hease, W. and Sejil, S. and Senellart, P. and Lemaître, A. and Ducci, S. and Leo, G. and Favero, I.},
	month = dec,
	year = {2013},
	pages = {241112},
}

@article{nguyen_improved_2015,
	title = {Improved optomechanical disk resonator sitting on a pedestal mechanical shield},
	volume = {17},
	issn = {1367-2630},
	url = {https://iopscience.iop.org/article/10.1088/1367-2630/17/2/023016},
	doi = {10.1088/1367-2630/17/2/023016},
	language = {english},
	number = {2},
	urldate = {2025-03-13},
	journal = {New Journal of Physics},
	author = {Nguyen, Dac Trung and Hease, William and Baker, Christopher and Gil-Santos, Eduardo and Senellart, Pascale and Lemaître, Aristide and Ducci, Sara and Leo, Giuseppe and Favero, Ivan},
	month = feb,
	year = {2015},
	pages = {023016},
}

@article{srivastava_identification_2023,
	title = {Identification of a {Large} {Number} of {Spin}-{Wave} {Eigenmodes} {Excited} by {Parametric} {Pumping} in {Yttrium} {Iron} {Garnet} {Microdisks}},
	volume = {19},
	url = {https://link.aps.org/doi/10.1103/PhysRevApplied.19.064078},
	doi = {10.1103/PhysRevApplied.19.064078},
	number = {6},
	urldate = {2025-02-11},
	journal = {Physical Review Applied},
	author = {Srivastava, T. and Merbouche, H. and Ngouagnia Yemeli, I. and Beaulieu, N. and Ben Youssef, J. and Muñoz, M. and Che, P. and Bortolotti, P. and Cros, V. and Klein, O. and Sangiao, S. and De Teresa, J.M. and Demokritov, S.O. and Demidov, V.E. and Anane, A. and Serpico, C. and d’Aquino, M. and de Loubens, G.},
	month = jun,
	year = {2023},
	note = {Publisher: American Physical Society},
	pages = {064078},
}

@article{hamadeh_core_2024,
	title = {Core {Reversal} in {Vertically} {Coupled} {Vortices}: {Simulation} and {Experimental} {Study}},
	volume = {23},
	issn = {1941-0085},
	shorttitle = {Core {Reversal} in {Vertically} {Coupled} {Vortices}},
	url = {https://ieeexplore.ieee.org/abstract/document/10577190},
	doi = {10.1109/TNANO.2024.3420249},
	urldate = {2025-02-11},
	journal = {IEEE Transactions on Nanotechnology},
	author = {Hamadeh, Abbass and Koujok, Abbas and Perna, Salvatore and Rodrigues, Davi R. and Riveros, Alejandro and Lomakin, Vitaliy and Finocchio, Giovanni and de Loubens, Grégoire and Klein, Olivier and Pirro, Philipp},
	year = {2024},
	note = {Conference Name: IEEE Transactions on Nanotechnology},
	pages = {549--553},
}

@misc{seeger_experimental_2024,
	title = {Experimental observation of vortex gyrotropic mode excited by surface acoustic waves},
	url = {http://arxiv.org/abs/2409.05998},
	doi = {10.48550/arXiv.2409.05998},
	urldate = {2025-02-11},
	publisher = {arXiv},
	author = {Seeger, R. Lopes and Millo, F. and Soares, G. and Kim, J.-V. and Solignac, A. and Loubens, G. de and Devolder, T.},
	month = oct,
	year = {2024},
	note = {arXiv:2409.05998 [cond-mat]},
}

@article{shinjo_magnetic_2000,
	title = {Magnetic {Vortex} {Core} {Observation} in {Circular} {Dots} of {Permalloy}},
	volume = {289},
	url = {https://www.science.org/doi/10.1126/science.289.5481.930},
	doi = {10.1126/science.289.5481.930},
	number = {5481},
	urldate = {2025-01-22},
	journal = {Science},
	author = {Shinjo, T. and Okuno, T. and Hassdorf, R. and Shigeto,  K. and Ono, T.},
	month = aug,
	year = {2000},
	note = {Publisher: American Association for the Advancement of Science},
	pages = {930--932},
}

@article{guslienko_magnetic_2008,
	title = {Magnetic {Vortex} {State} {Stability}, {Reversal} and {Dynamics} in {Restricted} {Geometries}},
	volume = {8},
	doi = {10.1166/jnn.2008.18305},
	number = {6},
	journal = {Journal of Nanoscience and Nanotechnology},
	author = {Guslienko, K. Yu.},
	month = jun,
	year = {2008},
	pages = {2745--2760},
}

@article{ding_wavelength-sized_2011,
	title = {Wavelength-sized {GaAs} optomechanical resonators with gigahertz frequency},
	volume = {98},
	issn = {0003-6951},
	url = {https://doi.org/10.1063/1.3563711},
	doi = {10.1063/1.3563711},
	number = {11},
	urldate = {2025-01-22},
	journal = {Applied Physics Letters},
	author = {Ding, L. and Baker, C. and Senellart, P. and Lemaitre, A. and Ducci, S. and Leo, G. and Favero, I.},
	month = mar,
	year = {2011},
	pages = {113108},
}

@article{matsko_optomechanics_2009,
	title = {Optomechanics with {Surface}-{Acoustic}-{Wave} {Whispering}-{Gallery} {Modes}},
	volume = {103},
	url = {https://link.aps.org/doi/10.1103/PhysRevLett.103.257403},
	doi = {10.1103/PhysRevLett.103.257403},
	number = {25},
	urldate = {2025-01-22},
	journal = {Physical Review Letters},
	author = {Matsko, A. B. and Savchenkov, A. A. and Ilchenko, V. S. and Seidel, D. and Maleki, L.},
	month = dec,
	year = {2009},
	note = {Publisher: American Physical Society},
	pages = {257403},
}

@article{savchenkov_surface_2011,
	title = {Surface acoustic wave opto-mechanical oscillator and frequency comb generator},
	volume = {36},
	copyright = {© 2011 Optical Society of America},
	issn = {1539-4794},
	url = {https://opg.optica.org/ol/abstract.cfm?uri=ol-36-17-3338},
	doi = {10.1364/OL.36.003338},
	language = {english},
	number = {17},
	urldate = {2025-01-22},
	journal = {Optics Letters},
	author = {Savchenkov, A. A. and Matsko, A. B. and Ilchenko, V. S. and Seidel, D. and Maleki, L.},
	month = sep,
	year = {2011},
	note = {Publisher: Optica Publishing Group},
	pages = {3338--3340},
}

@article{losby_torque-mixing_2015,
	title = {Torque-mixing magnetic resonance spectroscopy},
	volume = {350},
	url = {https://www.science.org/doi/10.1126/science.aad2449},
	doi = {10.1126/science.aad2449},
	number = {6262},
	urldate = {2025-01-22},
	journal = {Science},
	author = {Losby, J. E. and Sani, F. Fani and Grandmont, D. T. and Diao, Z. and Belov, M. and Burgess, J. A. J. and Compton, S. R. and Hiebert, W. K. and Vick, D. and Mohammad, K. and Salimi, E. and Bridges, G. E. and Thomson, D. J. and Freeman, M. R.},
	month = nov,
	year = {2015},
	note = {Publisher: American Association for the Advancement of Science},
	pages = {798--801},
}

@article{collet_generation_2016,
	title = {Generation of coherent spin-wave modes in yttrium iron garnet microdiscs by spin–orbit torque},
	volume = {7},
	copyright = {2016 The Author(s)},
	issn = {2041-1723},
	url = {https://www.nature.com/articles/ncomms10377},
	doi = {10.1038/ncomms10377},
	language = {english},
	number = {1},
	urldate = {2025-01-22},
	journal = {Nature Communications},
	author = {Collet, M. and de Milly, X. and d’Allivy Kelly, O. and Naletov, V. V. and Bernard, R. and Bortolotti, P. and Ben Youssef, J. and Demidov, V. E. and Demokritov, S. O. and Prieto, J. L. and Muñoz, M. and Cros, V. and Anane, A. and de Loubens, G. and Klein, O.},
	month = jan,
	year = {2016},
	note = {Publisher: Nature Publishing Group},
	pages = {10377},
}

@article{zhu_patterned_2017,
	title = {Patterned growth of crystalline {Y3Fe5O12} nanostructures with engineered magnetic shape anisotropy},
	volume = {110},
	issn = {0003-6951},
	url = {https://doi.org/10.1063/1.4986474},
	doi = {10.1063/1.4986474},
	number = {25},
	urldate = {2025-01-22},
	journal = {Applied Physics Letters},
	author = {Zhu, Na and Chang, Houchen and Franson, Andrew and Liu, Tao and Zhang, Xufeng and Johnston-Halperin, E. and Wu, Mingzhong and Tang, Hong X.},
	month = jun,
	year = {2017},
	pages = {252401},
}

@article{Lambert,
    author = {Lambert, Nicholas J. and Rueda, Alfredo and Sedlmeir, Florian and Schwefel, Harald G. L.},
    title = {Coherent Conversion Between Microwave and Optical Photons—An Overview of Physical Implementations},
    journal = {Advanced Quantum Technologies},
    volume = {3},
    number = {1},
    pages = {1900077},
    doi = {https://doi.org/10.1002/qute.201900077},
    url = {https://onlinelibrary.wiley.com/doi/abs/10.1002/qute.201900077},
    year = {2020}
}

@article{Lauk,
    doi = {10.1088/2058-9565/ab788a},
    url = {https://dx.doi.org/10.1088/2058-9565/ab788a},
    year = {2020},
    month = {mar},
    publisher = {IOP Publishing},
    volume = {5},
    number = {2},
    pages = {020501},
    author = {Nikolai Lauk and Neil Sinclair and Shabir Barzanjeh and Jacob P Covey and Mark Saffman and Maria Spiropulu and Christoph Simon},
    title = {Perspectives on quantum transduction},
    journal = {Quantum Science and Technology}
}

@article{Sahu,
    author={Sahu, Rishabh and Hease, William and Rueda, Alfredo and Arnold, Georg and Qiu, Liu and Fink, Johannes M.},
    title={Quantum-enabled operation of a microwave-optical interface},
    journal={Nature Communications},
    year={2022},
    month={Mar},
    day={11},
    volume={13},
    number={1},
    pages={1276},
    issn={2041-1723},
    doi={10.1038/s41467-022-28924-2},
    url={https://doi.org/10.1038/s41467-022-28924-2}
}

@Article{Bochmann,
    author={Bochmann, Joerg and Vainsencher, Amit and Awschalom, David D. and Cleland, Andrew N.},
    title={Nanomechanical coupling between microwave and optical photons},
    journal={Nature Physics},
    year={2013},
    month={Nov},
    day={01},
    volume={9},
    number={11},
    pages={712-716},
    issn={1745-2481},
    doi={10.1038/nphys2748},
    url={https://doi.org/10.1038/nphys2748}
}

@Article{Bagci,
    author={Bagci, T. and Simonsen, A. and Schmid, S. and Villanueva, L. G. and Zeuthen, E. and Appel, J. and Taylor, J. M. and S{\o}rensen, A. and Usami, K. and Schliesser, A. and Polzik, E. S.},
    title={Optical detection of radio waves through a nanomechanical transducer},
    journal={Nature},
    year={2014},
    month={Mar},
    day={01},
    volume={507},
    number={7490},
    pages={81-85},
    issn={1476-4687},
    doi={10.1038/nature13029},
    url={https://doi.org/10.1038/nature13029}
}

@Article{Regal2014,
    author={Andrews, R. W. and Peterson, R. W. and Purdy, T. P. and Cicak, K. and Simmonds, R. W. and Regal, C. A. and Lehnert, K. W.},
    title={Bidirectional and efficient conversion between microwave and optical light},
    journal={Nature Physics},
    year={2014},
    month={Apr},
    day={01},
    volume={10},
    number={4},
    pages={321-326},
    issn={1745-2481},
    doi={10.1038/nphys2911},
    url={https://doi.org/10.1038/nphys2911}
}

@Article{Forsch,
    author={Forsch, Moritz and Stockill, Robert and Wallucks, Andreas and Marinkovi{\'{c}}, Igor and G{\"a}rtner, Claus and Norte, Richard A. and van Otten, Frank and Fiore, Andrea and Srinivasan, Kartik and Gr{\"o}blacher, Simon},
    title={Microwave-to-optics conversion using a mechanical oscillator in its quantum ground state},
    journal={Nature Physics},
    year={2020},
    month={Jan},
    day={01},
    volume={16},
    number={1},
    pages={69-74},
    issn={1745-2481},
    doi={10.1038/s41567-019-0673-7},
    url={https://doi.org/10.1038/s41567-019-0673-7}
}

@article{Hisatomi,
    title = {Bidirectional conversion between microwave and light via ferromagnetic magnons},
    author = {Hisatomi, R. and Osada, A. and Tabuchi, Y. and Ishikawa, T. and Noguchi, A. and Yamazaki, R. and Usami, K. and Nakamura, Y.},
    journal = {Phys. Rev. B},
    volume = {93},
    issue = {17},
    pages = {174427},
    numpages = {13},
    year = {2016},
    month = {May},
    publisher = {American Physical Society},
    doi = {10.1103/PhysRevB.93.174427},
    url = {https://link.aps.org/doi/10.1103/PhysRevB.93.174427}
}

@article{Zhu2020,
    author = {Na Zhu and Xufeng Zhang and Xu Han and Chang-Ling Zou and Changchun Zhong and Chiao-Hsuan Wang and Liang Jiang and Hong X. Tang},
    journal = {Optica},
    number = {10},
    pages = {1291--1297},
    publisher = {Optica Publishing Group},
    title = {Waveguide cavity optomagnonics for microwave-to-optics conversion},
    volume = {7},
    month = {Oct},
    year = {2020},
    url = {https://opg.optica.org/optica/abstract.cfm?URI=optica-7-10-1291},
    doi = {10.1364/OPTICA.397967},
}

@Article{Regal2018,
    author={Higginbotham, A. P. and Burns, P. S. and Urmey, M. D. and Peterson, R. W. and Kampel, N. S. and Brubaker, B. M. and Smith, G. and Lehnert, K. W. and Regal, C. A.},
    title={Harnessing electro-optic correlations in an efficient mechanical converter},
    journal={Nature Physics},
    year={2018},
    month={Oct},
    day={01},
    volume={14},
    number={10},
    pages={1038-1042},
    issn={1745-2481},
    doi={10.1038/s41567-018-0210-0},
    url={https://doi.org/10.1038/s41567-018-0210-0}
}

@article{Rameshti,
    title = {Cavity magnonics},
    journal = {Physics Reports},
    volume = {979},
    pages = {1-61},
    year = {2022},
    note = {Cavity Magnonics},
    issn = {0370-1573},
    doi = {https://doi.org/10.1016/j.physrep.2022.06.001},
    url = {https://www.sciencedirect.com/science/article/pii/S0370157322002460},
    author = {Babak {Zare Rameshti} and Silvia {Viola Kusminskiy} and James A. Haigh and Koji Usami and Dany Lachance-Quirion and Yasunobu Nakamura and Can-Ming Hu and Hong X. Tang and Gerrit E. W. Bauer and Yaroslav M. Blanter}
}

@article{Engelhardt,
    title = {Optimal Broadband Frequency Conversion via a Magnetomechanical Transducer},
    author = {Engelhardt, F. and Bittencourt, V.A.S.V. and Huebl, H. and Klein, O. and Kusminskiy, S. Viola},
    journal = {Phys. Rev. Appl.},
    volume = {18},
    issue = {4},
    pages = {044059},
    numpages = {16},
    year = {2022},
    month = {Oct},
    publisher = {American Physical Society},
    doi = {10.1103/PhysRevApplied.18.044059},
    url = {https://link.aps.org/doi/10.1103/PhysRevApplied.18.044059}
}

@article{An2020,
    title = {Coherent long-range transfer of angular momentum between magnon Kittel modes by phonons},
    author = {An, K. and Litvinenko, A. N. and Kohno, R. and Fuad, A. A. and Naletov, V. V. and Vila, L. and Ebels, U. and de Loubens, G. and Hurdequint, H. and Beaulieu, N. and Ben Youssef, J. and Vukadinovic, N. and Bauer, G. E. W. and Slavin, A. N. and Tiberkevich, V. S. and Klein, O.},
    journal = {Phys. Rev. B},
    volume = {101},
    issue = {6},
    pages = {060407},
    numpages = {6},
    year = {2020},
    month = {Feb},
    publisher = {American Physical Society},
    doi = {10.1103/PhysRevB.101.060407},
    url = {https://link.aps.org/doi/10.1103/PhysRevB.101.060407}
}

@article{Schlitz,
    title = {Magnetization dynamics affected by phonon pumping},
    author = {Schlitz, Richard and Siegl, Luise and Sato, Takuma and Yu, Weichao and Bauer, Gerrit E. W. and Huebl, Hans and Goennenwein, Sebastian T. B.},
    journal = {Phys. Rev. B},
    volume = {106},
    issue = {1},
    pages = {014407},
    numpages = {10},
    year = {2022},
    month = {Jul},
    publisher = {American Physical Society},
    doi = {10.1103/PhysRevB.106.014407},
    url = {https://link.aps.org/doi/10.1103/PhysRevB.106.014407}
}

@article{Hatanaka,
    title = {On-Chip Coherent Transduction between Magnons and Acoustic Phonons in Cavity Magnomechanics},
    author = {Hatanaka, D. and Asano, M. and Okamoto, H. and Kunihashi, Y. and Sanada, H. and Yamaguchi, H.},
    journal = {Phys. Rev. Appl.},
    volume = {17},
    issue = {3},
    pages = {034024},
    numpages = {12},
    year = {2022},
    month = {Mar},
    publisher = {American Physical Society},
    doi = {10.1103/PhysRevApplied.17.034024},
    url = {https://link.aps.org/doi/10.1103/PhysRevApplied.17.034024}
}

@article{Schmidt1,
    author = {Schmidt, Georg and Hauser, Christoph and Trempler, Philip and Paleschke, Maximilian and Papaioannou, Evangelos Th.},
    title = {Ultra Thin Films of Yttrium Iron Garnet with Very Low Damping: A Review},
    journal = {physica status solidi (b)},
    volume = {257},
    number = {7},
    pages = {1900644},
    doi = {https://doi.org/10.1002/pssb.201900644},
    url = {https://onlinelibrary.wiley.com/doi/abs/10.1002/pssb.201900644},
    eprint = {https://onlinelibrary.wiley.com/doi/pdf/10.1002/pssb.201900644},
    year = {2020}
}

@article{Schmidt2,
    author = {Trempler, P. and Dreyer, R. and Geyer, P. and Hauser, C. and Woltersdorf, G. and Schmidt, G.},
    title = "{Integration and characterization of micron-sized YIG structures with very low Gilbert damping on arbitrary substrates}",
    journal = {Applied Physics Letters},
    volume = {117},
    number = {23},
    pages = {232401},
    year = {2020},
    month = {12},
    issn = {0003-6951},
    doi = {10.1063/5.0026120},
    url = {https://doi.org/10.1063/5.0026120},
    eprint = {https://pubs.aip.org/aip/apl/article-pdf/doi/10.1063/5.0026120/14541836/232401\_1\_online.pdf},
}

@article{Graf2018,
  title = {Cavity optomagnonics with magnetic textures: Coupling a magnetic vortex to light},
  author = {Graf, Jasmin and Pfeifer, Hannes and Marquardt, Florian and Viola Kusminskiy, Silvia},
  journal = {Phys. Rev. B},
  volume = {98},
  issue = {24},
  pages = {241406},
  numpages = {5},
  year = {2018},
  month = {Dec},
  publisher = {American Physical Society},
  doi = {10.1103/PhysRevB.98.241406},
  url = {https://link.aps.org/doi/10.1103/PhysRevB.98.241406}
}

@article{Clark,
    author = {Clark, A. E. and Strakna, R. E.},
    title = "{Elastic Constants of Single‐Crystal YIG}",
    journal = {Journal of Applied Physics},
    volume = {32},
    number = {6},
    pages = {1172-1173},
    year = {1961},
    month = {06},
    issn = {0021-8979},
    doi = {10.1063/1.1736184},
    url = {https://doi.org/10.1063/1.1736184},
    eprint = {https://pubs.aip.org/aip/jap/article-pdf/32/6/1172/18323984/1172\_1\_online.pdf},
}

@article{Vansteenkiste,
    author = {Vansteenkiste, Arne and Leliaert, Jonathan and Dvornik, Mykola and Helsen, Mathias and Garcia-Sanchez, Felipe and Van Waeyenberge, Bartel},
    title = "{The design and verification of MuMax3}",
    journal = {AIP Advances},
    volume = {4},
    number = {10},
    pages = {107133},
    year = {2014},
    month = {10},
    issn = {2158-3226},
    doi = {10.1063/1.4899186},
    url = {https://doi.org/10.1063/1.4899186},
    eprint = {https://pubs.aip.org/aip/adv/article-pdf/doi/10.1063/1.4899186/12878560/107133\_1\_online.pdf},
}

@article{Kakazei,
    author = {Kakazei, G. N. and Wigen, P. E. and Guslienko, K. Yu. and Novosad, V. and Slavin, A. N. and Golub, V. O. and Lesnik, N. A. and Otani, Y.},
    title = "{Spin-wave spectra of perpendicularly magnetized circular submicron dot arrays}",
    journal = {Applied Physics Letters},
    volume = {85},
    number = {3},
    pages = {443-445},
    year = {2004},
    month = {07},
    issn = {0003-6951},
    doi = {10.1063/1.1772868},
    url = {https://doi.org/10.1063/1.1772868},
    eprint = {https://pubs.aip.org/aip/apl/article-pdf/85/3/443/18591770/443\_1\_online.pdf},
}

@Article{Dobtrovolskiy,
author ="Dobrovolskiy, Oleksandr V. and Bunyaev, Sergey A. and Vovk, Nikolay R. and Navas, David and Gruszecki, Pawel and Krawczyk, Maciej and Sachser, Roland and Huth, Michael and Chumak, Andrii V. and Guslienko, Konstantin Y. and Kakazei, Gleb N.",
title  ="Spin-wave spectroscopy of individual ferromagnetic nanodisks",
journal  ="Nanoscale",
year  ="2020",
volume  ="12",
issue  ="41",
pages  ="21207-21217",
publisher  ="The Royal Society of Chemistry",
doi  ="10.1039/D0NR07015G",
url  ="http://dx.doi.org/10.1039/D0NR07015G"}

@article{Ivanov2005,
    author = {Ivanov, B. A.},
    title = "{Mesoscopic antiferromagnets: statics, dynamics, and quantum tunneling (Review)}",
    journal = {Low Temperature Physics},
    volume = {31},
    number = {8},
    pages = {635-667},
    year = {2005},
    month = {08},
    issn = {1063-777X},
    doi = {10.1063/1.2008127},
    url = {https://doi.org/10.1063/1.2008127},
    eprint = {https://pubs.aip.org/aip/ltp/article-pdf/31/8/635/8232512/635\_1\_online.pdf},
}

@article{Radcliffe,
    doi = {10.1088/0305-4470/4/3/009},
    url = {https://dx.doi.org/10.1088/0305-4470/4/3/009},
    year = {1971},
    month = {may},
    publisher = {},
    volume = {4},
    number = {3},title = {Helicity of magnetic vortices and skyrmions in soft ferromagnetic nanodots and films biased by stray radial fields},
author = {Verba, R. V. and Navas, D. and Bunyaev, S. A. and Hierro-Rodriguez, A. and Guslienko, K. Y. and Ivanov, B. A. and Kakazei, G. N.},
journal = {Phys. Rev. B},
volume = {101},
issue = {6},
pages = {064429},
numpages = {13},
year = {2020},
month = {Feb},
publisher = {American Physical Society},
doi = {10.1103/PhysRevB.101.064429},
url = {https://link.aps.org/doi/10.1103/PhysRevB.101.064429},
    pages = {313},
    author = {J M Radcliffe},
    title = {Some properties of coherent spin states},
    journal = {Journal of Physics A: General Physics}
}

@Article{Feldtkeller,
    author={Feldtkeller, Ernst  and Thomas, Harry},
    title={Struktur und Energie von Blochlinien in d{\"u}nnen ferromagnetischen Schichten},
    journal={Physik der kondensierten Materie},
    year={1965},
    month={Jul},
    day={01},
    volume={4},
    number={1},
    pages={8-14},
    issn={1431-584X},
    doi={10.1007/BF02423256},
    url={https://doi.org/10.1007/BF02423256}
}

@article{Clark1963,
    author = {Clark, A. E. and DeSavage, B. and Coleman, W. and Callen, E. R. and Callen, H. B.},
    title = "{Saturation Magnetostriction of Single‐Crystal YIG}",
    journal = {Journal of Applied Physics},
    volume = {34},
    number = {4},
    pages = {1296-1297},
    year = {1963},
    month = {04},
    issn = {0021-8979},
    doi = {10.1063/1.1729480},
    url = {https://doi.org/10.1063/1.1729480},
    eprint = {https://pubs.aip.org/aip/jap/article-pdf/34/4/1296/18328747/1296\_1\_online.pdf},
}

@misc{Platzker,
    author = {Platzker, Aryeh},
    title = "{MAGNETOELASTIC PROPERTIES OF YIG AS A FUNCTION OF TEMPERATURE}",
    howpublished = {AD0655085},
    year = {1967},
    url = {https://apps.dtic.mil/sti/citations/AD0655085}
}

@article{Verba2020,
    title = {Helicity of magnetic vortices and skyrmions in soft ferromagnetic nanodots and films biased by stray radial fields},
    author = {Verba, R. V. and Navas, D. and Bunyaev, S. A. and Hierro-Rodriguez, A. and Guslienko, K. Y. and Ivanov, B. A. and Kakazei, G. N.},
    journal = {Phys. Rev. B},
    volume = {101},
    issue = {6},
    pages = {064429},
    numpages = {13},
    year = {2020},
    month = {Feb},
    publisher = {American Physical Society},
    doi = {10.1103/PhysRevB.101.064429},
    url = {https://link.aps.org/doi/10.1103/PhysRevB.101.064429}
}

@article{PhysRev.110.836,
  title = {Interaction of Spin Waves and Ultrasonic Waves in Ferromagnetic Crystals},
  author = {Kittel, C.},
  journal = {Phys. Rev.},
  volume = {110},
  issue = {4},
  pages = {836--841},
  numpages = {0},
  year = {1958},
  month = {May},
  publisher = {American Physical Society},
  doi = {10.1103/PhysRev.110.836},
  url = {https://link.aps.org/doi/10.1103/PhysRev.110.836}
}

@article{Aspelmeyer_2014,
   title={Cavity optomechanics},
   volume={86},
   ISSN={1539-0756},
   url={http://dx.doi.org/10.1103/RevModPhys.86.1391},
   DOI={10.1103/revmodphys.86.1391},
   number={4},
   journal={Reviews of Modern Physics},
   publisher={American Physical Society (APS)},
   author={Aspelmeyer, Markus and Kippenberg, Tobias J. and Marquardt, Florian},
   year={2014},
   month=dec, pages={1391–1452} }

@article{
doi:10.1126/science.1075302,
author = {A. Wachowiak  and J. Wiebe  and M. Bode  and O. Pietzsch  and M. Morgenstern  and R. Wiesendanger },
title = {Direct Observation of Internal Spin Structure of Magnetic Vortex Cores},
journal = {Science},
volume = {298},
number = {5593},
pages = {577-580},
year = {2002},
doi = {10.1126/science.1075302},
URL = {https://www.science.org/doi/abs/10.1126/science.1075302},
eprint = {https://www.science.org/doi/pdf/10.1126/science.1075302}}

@article{PhysRevLett.94.027205,
  title = {High Frequency Modes in Vortex-State Nanomagnets},
  author = {Ivanov, B. A. and Zaspel, C. E.},
  journal = {Phys. Rev. Lett.},
  volume = {94},
  issue = {2},
  pages = {027205},
  numpages = {4},
  year = {2005},
  month = {Jan},
  publisher = {American Physical Society},
  doi = {10.1103/PhysRevLett.94.027205},
  url = {https://link.aps.org/doi/10.1103/PhysRevLett.94.027205}
}

@Article{D4NH00145A,
author ="Bondarenko, Artem V. and Bunyaev, Sergey A. and Shukla, Amit K. and Apolinario, Arlete and Singh, Navab and Navas, David and Guslienko, Konstantin Y. and Adeyeye, Adekunle O. and Kakazei, Gleb N.",
title  ="Dominant higher-order vortex gyromodes in circular magnetic nanodots",
journal  ="Nanoscale Horiz.",
year  ="2024",
volume  ="9",
issue  ="9",
pages  ="1498-1505",
publisher  ="The Royal Society of Chemistry",
doi  ="10.1039/D4NH00145A",
url  ="http://dx.doi.org/10.1039/D4NH00145A"}

@article{10.1063/1.1652420,
    author = {Ivanov, B. A. and Zaspel, C. E.},
    title = "{Gyrotropic mode frequency of vortex-state permalloy disks}",
    journal = {Journal of Applied Physics},
    volume = {95},
    number = {11},
    pages = {7444-7446},
    year = {2004},
    month = {06},
    issn = {0021-8979},
    doi = {10.1063/1.1652420},
    url = {https://doi.org/10.1063/1.1652420},
    eprint = {https://pubs.aip.org/aip/jap/article-pdf/95/11/7444/18710283/7444\_1\_online.pdf},
}

@article{PhysRevB.58.8464,
  title = {Magnon modes and magnon-vortex scattering in two-dimensional easy-plane ferromagnets},
  author = {Ivanov, B. A. and Schnitzer, H. J. and Mertens, F. G. and Wysin, G. M.},
  journal = {Phys. Rev. B},
  volume = {58},
  issue = {13},
  pages = {8464--8474},
  numpages = {0},
  year = {1998},
  month = {Oct},
  publisher = {American Physical Society},
  doi = {10.1103/PhysRevB.58.8464},
  url = {https://link.aps.org/doi/10.1103/PhysRevB.58.8464}
}

@Article{Stockill2022,
author={Stockill, Robert
and Forsch, Moritz
and Hijazi, Frederick
and Beaudoin, Gr{\~A}{\textcopyright}goire
and Pantzas, Konstantinos
and Sagnes, Isabelle
and Braive, R{\~A}{\textcopyright}my
and Gr{\"o}blacher, Simon},
title={Ultra-low-noise microwave to optics conversion in gallium phosphide},
journal={Nature Communications},
year={2022},
month={Nov},
day={03},
volume={13},
number={1},
pages={6583},
issn={2041-1723},
doi={10.1038/s41467-022-34338-x},
url={https://doi.org/10.1038/s41467-022-34338-x}
}

@book{makarov2022curvilinear,
  title={Curvilinear Micromagnetism: From Fundamentals to Applications},
  author={Makarov, D. and Sheka, D.D.},
  isbn={9783031090851},
  series={Topics in Applied Physics},
  url={https://books.google.nl/books?id=3MMszwEACAAJ},
  year={2022},
  publisher={Springer International Publishing}
}

@book{Landau1986,
  asin = {075062633X},
  author = {Landau, L. D. and Pitaevskii, L. P. and Lifshitz, E. M. and Kosevich, A. M.},
  chapter = {25},
  dewey = {531.38},
  ean = {9780750626330},
  edition = 3,
  isbn = {075062633X},
  publisher = {Butterworth-Heinemann},
  title = {Theory of Elasticity},
  url = {http://www.amazon.com/Theory-Elasticity-Third-Theoretical-Physics/dp/075062633X/ref=sr_1_16?ie=UTF8&s=books&qid=1280929419&sr=8-16},
  year = 1986
}

@book{Mavko_Mukerji_Dvorkin_2009,
    place={Cambridge},
    edition={2},
    title={The Rock Physics Handbook: Tools for Seismic Analysis of Porous Media},
    publisher={Cambridge University Press},
    chapter={2.2},
    author={Mavko, Gary and Mukerji, Tapan and Dvorkin, Jack},
    year={2009}
}

@incollection{SUN2013157,
title = {Chapter Six - Yttrium Iron Garnet Nano Films: Epitaxial Growth, Spin-Pumping Efficiency, and Pt-Capping-Caused Damping},
editor = {Mingzhong Wu and Axel Hoffmann},
series = {Solid State Physics},
publisher = {Academic Press},
volume = {64},
pages = {157-191},
year = {2013},
booktitle = {Recent Advances in Magnetic Insulators – From Spintronics to Microwave Applications},
issn = {0081-1947},
doi = {https://doi.org/10.1016/B978-0-12-408130-7.00006-X},
url = {https://www.sciencedirect.com/science/article/pii/B978012408130700006X},
author = {Yiyan Sun and Mingzhong Wu}
}

@book{auld1973acoustic,
  title={Acoustic fields and waves in solids},
  author={Auld, B.A.},
  isbn={9785885013437},
  series={A Wiley-Interscience publication},
  url={https://books.google.nl/books?id=_2MWAwAAQBAJ},
  year={1973},
  publisher={Wiley}
}

@article{PhysRevB.76.014416,
  title = {Mode degeneracy due to vortex core removal in magnetic disks},
  author = {Hoffmann, F. and Woltersdorf, G. and Perzlmaier, K. and Slavin, A. N. and Tiberkevich, V. S. and Bischof, A. and Weiss, D. and Back, C. H.},
  journal = {Phys. Rev. B},
  volume = {76},
  issue = {1},
  pages = {014416},
  numpages = {5},
  year = {2007},
  month = {Jul},
  publisher = {American Physical Society},
  doi = {10.1103/PhysRevB.76.014416},
  url = {https://link.aps.org/doi/10.1103/PhysRevB.76.014416}
}

@article{PhysRevB.90.064410,
  title = {Breathing modes of confined skyrmions in ultrathin magnetic dots},
  author = {Kim, Joo-Von and Garcia-Sanchez, Felipe and Sampaio, Jo\~ao and Moreau-Luchaire, Constance and Cros, Vincent and Fert, Albert},
  journal = {Phys. Rev. B},
  volume = {90},
  issue = {6},
  pages = {064410},
  numpages = {8},
  year = {2014},
  month = {Aug},
  publisher = {American Physical Society},
  doi = {10.1103/PhysRevB.90.064410},
  url = {https://link.aps.org/doi/10.1103/PhysRevB.90.064410}
}

@article{PhysRevB.65.134434,
  title = {Magnon modes for a circular two-dimensional easy-plane ferromagnet in the cone state},
  author = {Ivanov, B. A. and Wysin, G. M.},
  journal = {Phys. Rev. B},
  volume = {65},
  issue = {13},
  pages = {134434},
  numpages = {17},
  year = {2002},
  month = {Mar},
  publisher = {American Physical Society},
  doi = {10.1103/PhysRevB.65.134434},
  url = {https://link.aps.org/doi/10.1103/PhysRevB.65.134434}
}

@article{Heyroth2019,
  title = {Monocrystalline Freestanding Three-Dimensional Yttrium-Iron-Garnet Magnon Nanoresonators},
  volume = {12},
  ISSN = {2331-7019},
  url = {http://dx.doi.org/10.1103/PhysRevApplied.12.054031},
  DOI = {10.1103/physrevapplied.12.054031},
  number = {5},
  journal = {Physical Review Applied},
  publisher = {American Physical Society (APS)},
  author = {Heyroth,  F. and Hauser,  C. and Trempler,  P. and Geyer,  P. and Syrowatka,  F. and Dreyer,  R. and Ebbinghaus,  S.G. and Woltersdorf,  G. and Schmidt,  G.},
  year = {2019},
  month = nov 
}

@Inbook{Vanderveken2021,
author="Vanderveken, Frederic
and Ciubotaru, Florin
and Adelmann, Christoph",
editor="Kamenetskii, Eugene",
title="Magnetoelastic Waves in Thin Films",
bookTitle="Chirality, Magnetism and Magnetoelectricity: Separate Phenomena and Joint Effects in Metamaterial Structures",
year="2021",
publisher="Springer International Publishing",
address="Cham",
pages="287--322",
isbn="978-3-030-62844-4",
doi="10.1007/978-3-030-62844-4_12",
url="https://doi.org/10.1007/978-3-030-62844-4_12"
}

\appendix
\section{\label{app:tens_der}4-rank tensor transformations and cylindrical coordinate tensors}

In this manuscript, while considering magnetoelastic interaction, we have to deal with 4-rank tensors. Although it is possible to represent such an object in its full complexity, the fundamental symmetries of the problem reduce the 4-rank tensor $A_{ijkl}$ to only $36$ unique elements in the Voigt notation (for more details see for example Refs.~\cite{Mavko_Mukerji_Dvorkin_2009,auld1973acoustic}), 
\begin{table}[h!]
\begin{tabular}{l|c c c c c c}
        Voigt index&1&2&3&4&5&6\\
        Coord pair&xx&yy&zz&yz&xz&xy
\end{tabular}
\end{table}
The 4-rank tensor is now represented as a matrix, and also symmetric 2-rank tensors can now be made into vectors,
\begin{equation}
    A_{ij}=\begin{pmatrix}A_{xx}&A_{xy}&A_{xz}\\
                          A_{xy}&A_{yy}&A_{yz}\\
                          A_{xz}&A_{yz}&A_{zz}\end{pmatrix}\Rightarrow
    \widetilde{A}_\rho=\begin{pmatrix}A_{xx}\\A_{yy}\\A_{zz}\\2A_{yz}\\2A_{xz}\\2A_{xy}\end{pmatrix} \ ,
\end{equation}
preserving the conventional matrix multiplication, i.e.
\begin{equation}
    C_{ijkl}\epsilon_{ij}\epsilon_{kl}\equiv \bm{\epsilon}\widetilde{C}\bm{\epsilon}.
\end{equation}
In this Appendix we represent our results in the Voigt notation even though underlying calculations for transformations are done with complete 4-tensors.

For material parameters like $C$ and $B$, all transformations are done locally. To do so, we first identify a transform of a vector into the new coordinate system which is given by the matrix $A:\mathbf{x}'=A\mathbf{x}$. For the relevant transformations, which are all rotations because we go from an orthonormal basis to another orthonormal basis, the matrix $A$ is self-adjoint, $A^{-1} = A^T$. Using this property it is straightforward to show that the underlying transform of a 4-rank tensor is given by the following expression,
\begin{equation}
    C'_{ijkl} = A_{i\alpha} A_{j\beta} A_{k\gamma} A_{l\delta} C_{\alpha \beta \gamma \delta},
\end{equation}
with the summation over repeated Greek letter indices.

For the YIG film in (100) orientation, the Voigt representation of the magnetostrictive tensor $B$ is given as follows~\cite{PhysRev.110.836},
\begin{equation}
    \widetilde{B}=\begin{pmatrix}
        \frac{2B_1}{3}&-\frac{B_1}{3}&-\frac{B_1}{3}&0&0&0\\
        -\frac{B_1}{3}&\frac{2B_1}{3}&-\frac{B_1}{3}&0&0&0\\
        -\frac{B_1}{3}&-\frac{B_1}{3}&\frac{2B_1}{3}&0&0&0\\
        0&0&0&\frac{B_2}{2}&0&0\\
        0&0&0&0&\frac{B_2}{2}&0\\
        0&0&0&0&0&\frac{B_2}{2}\\
    \end{pmatrix},
\end{equation}
which provides the magnetostrictive energy equivalent to the one shown previously\eqref{eq:100ms_energy} with the additional term $-1/3 B_1 \Tr \epsilon$. Transforming the matrix into (111) crystaline orientation is done by applying the following rotation,
\begin{equation}
    R_{(111)}=\begin{pmatrix}
        \frac{1}{\sqrt{2}} & \frac{1}{\sqrt{6}} & \frac{1}{\sqrt{3}} \\
        -\frac{1}{\sqrt{2}} & \frac{1}{\sqrt{6}} & \frac{1}{\sqrt{3}} \\
        0 & -\sqrt{\frac{2}{3}} & \frac{1}{\sqrt{3}}
    \end{pmatrix},
\end{equation}
which results in 
\begin{widetext}
\begin{equation}
    \centering
    \widetilde{B}_{(111)}=\begin{pmatrix}

        \frac{B_1+3 B_2}{6} & -\frac{B_1+B_2}{6} & -\frac{B_2}{3} & \frac{B_1-B_2}{3 \sqrt{2}} & 0 & 0 \\
         -\frac{B_1+B_2}{6} & \frac{B_1+3 B_2}{6} & -\frac{B_2}{3} & -\frac{B_1-B_2}{3 \sqrt{2}} & 0 & 0 \\
         -\frac{B_2}{3} & -\frac{B_2}{3} & \frac{2B_2}{3} & 0 & 0 & 0 \\
        \frac{B_1-B_2}{3 \sqrt{2}} & -\frac{B_1-B_2}{3 \sqrt{2}} & 0 & \frac{2 B_1+B_2}{6} & 0 & 0 \\
        0 & 0 & 0 & 0 & \frac{2 B_1+B_2}{6} & \frac{B_1-B_2}{3 \sqrt{2}} \\
        0 & 0 & 0 & 0 & \frac{B_1-B_2}{3 \sqrt{2}} & \frac{B_1+2B_2}{6} \\

    \end{pmatrix}.
\end{equation}
\end{widetext}

Next, in order to go to the cylindrical coordinate system, a rotation in plane is applied,
\begin{equation}
    R_{\varphi}=\begin{pmatrix}
        \cos\varphi &-\sin\varphi&0\\
        \sin\varphi & \cos\varphi&0\\
        0&0&1
    \end{pmatrix}.
\end{equation}
This transformation leads to a tensor which explicitly includes the anisotropy since its coefficients are functions of the azimuthal angle $\varphi$. Since we are either dealing with modes which have the azimuthal number $m=0$, or are interested in coupling terms of the form $m^\dagger b$, which as a result do not depend on $\varphi$, we can average the transformed tensor,
\begin{align}
    E_{ms}=&\int w_{ms} dV = \int r dr \int d\varphi \int dz m_i m_j \epsilon_{kl} B_{ijkl}=\notag\\
          =&2\pi\int r dr\int dz m_i m_j \epsilon_{kl} \left[\frac{1}{2\pi}\int_0^{2\pi}d\varphi B_{ijkl}\right].
\end{align}
The resulting averaged tensor for (111) film oriented with its surface normal to the axial z coordinate is as follows,
\begin{widetext}
\begin{equation}
    \widetilde{B}_{(111)}^\text{cyl.}= \begin{pmatrix}

        \frac{B_1+3 B_2}{6} & -\frac{B_1+B_2}{6} & -\frac{B_2}{3} & 0 & 0 & 0 \\
        -\frac{B_1+B_2}{6} & \frac{B_1+3 B_2}{6} & -\frac{B_2}{3} & 0 & 0 & 0 \\
        -\frac{B_2}{3} & -\frac{B_2}{3} & \frac{2 B_2}{3} & 0 & 0 & 0 \\
        0 & 0 & 0 & \frac{2 B_1+B_2}{6} & 0 & 0 \\
        0 & 0 & 0 & 0 & \frac{2 B_1+B_2}{6} & 0 \\
        0 & 0 & 0 & 0 & 0 & \frac{B_1+2B_2}{6} \\

    \end{pmatrix},
\end{equation}
\end{widetext}
with the column/row index Voigt elements corresponding to pairs $rr, \varphi\varphi, zz, \varphi z, rz, r\varphi$.

\section{Magnetic mode normalization}
\label{ap:magnon_norm}
The question of normalizing the magnetic modes is a non-trivial one, especially since the constitutive equations contain non-local dipolar interactions. The Lagrangian density itself in this case is a functional,
\begin{equation}
    \mathcal{L}_{mag.}[\bm{m}]=\hbar S \frac{\left[\bm{m}\times \mathbf{n}\right]}{m\left(m+\bm{m}\mathbf{\mathbf{n}}\right)}\dot{\bm{m}}-U[\bm{m}],
\end{equation}
where $\mathbf{n}$ is the quantization direction of spins, and $\bm{m}:\bm{m}^2=1$ is the dimensionless unit vector parameterizing the spin-coherent state angular momentum (eventually magnetization) orientation~\cite{Radcliffe,Ivanov2005}. The conjugate momentum $\bm{\pi}$ is easily derived from the expression, and it can be shown that the related Hamiltonian $\mathcal{H}=\bm{\pi}\bm{m}-\mathcal{L}=U$.

Next we consider that in the same way as with conventional derivatives, for a homogeneous functional $\mathcal{F}[s \mathbf{x}] = s^k \mathcal{F}[\mathbf{x}]$, we can take the $\partial_s$ derivative at $s=1$ to derive the Euler's identity analog for variational derivatives,
\begin{equation}
    k\mathcal{F}[\mathbf{x}]=x_i\frac{\delta \mathcal{F}}{\delta x_i}.
\end{equation}
After the problem is linearized near the stationary magnetization $\bm{m}_0$, only terms quadratic in $m$ remain in the resulting potential $U_{lin}$, and we can write the following dynamic equation,
\begin{equation}
    \dot{\pi_i}=-\frac{\delta\mathcal{H}}{\delta m_i},
\end{equation}
which, after being multiplied by $m_i$, gives the equality
\begin{equation}
    m_i\dot{\pi}_i = -m_i\frac{\delta U_{lin}}{\delta m_i} = -2 U_{lin} = -2\mathcal{H}_{lin},
\end{equation}
so that if we obtain the dynamic solution by other means, we can then calculate the energy of oscillations.

Next, we can use orthogonality of the conjugate momentum in the system  $\bm{m}\bm{\pi}\sim\left[\bm{m}\times \mathbf{n}\right]\bm{m} = 0$ to simplify the expression for the linearized Hamiltonian using the identity $\bm{m}\dot{\bm{\pi}}=\partial_t(\bm{m}\bm{\pi})-\dot{\bm{m}}\bm{\pi}=-\dot{\bm{m}}\bm{\pi}$,
\begin{equation}
    \mathcal{H}=\dot{\bm{m}}\bm{\pi}/2 = \frac{\hbar S}{2} \frac{\left[\bm{m}\times \mathbf{n}\right]}{m\left(m+\bm{m}\mathbf{\mathbf{n}}\right)}\dot{\bm{m}}.
\end{equation}
If we place the vector normal $\bm{n}=\bm{m}_0$ locally at every point $\bm{r}$, the scalar product in the denominator $\bm{m}\bm{m}_0$ is arbitrarily small. If we next assume that a solution was found externally such that $m=\bm{m}_0 + {\rm Re}\left\{\bm{a} e^{-i\omega t}\right\}$, and $\bm{a}=\bm{a}'+i \bm{a}''$, the expression for energy further simplifies to
\begin{equation}
    \mathcal{H}=\hbar \omega S \left[\bm{a}'' \times \bm{m}_0\right]\bm{a}',
\end{equation}
which gives a much easier way of computing the normalization constant without calculating the potential $U$ explicitly. Such calculation would be an additional operation, since solving the dynamic equations only evaluates the functional gradients of $U$. 

Finally, to switch to energy densities, and continuous medium magnetization $M_s\bm{m}$ we note that model constants can be replaced with the empirical ones after averaging the Hamiltonian over the elementary cell,
\begin{equation}
    \mathcal{H} = \omega \frac{M_s}{\gamma_0} \bm{m}_0\left[{\rm Im} (\bm{m}_d) \times {\rm Re} (\bm{m}_d)\right],
    \label{eq:mag_norm}
\end{equation}
where $M_s$ is saturation magnetization, and $\gamma_0$ is the gyromagnetic ratio.

The form of the expression repeats the one which was calculated earlier using the perturbation theory for the dynamic equations, and the one from the arguments presented for a specific structure of the potential $U$ earlier~\cite{Graf2018,Engelhardt}.
However, in contrast to other literature here we make absolute minimal assumptions, and, in principle, this approach is extendable to any linearized dynamic system with a known kinetic part of the Lagrangian, like elastic or electromagnetic field. This normalization has been checked against the direct numerical evaluation of $U$ in Mumax3.

\section{Elastic mode normalization}
\label{ap:phonon_norm}
Normalization of elastic modes is simplified further by the fact that the mean value of the total energy is double the mean of the kinetic energy $E=2T$, the property known as a virial theorem. The calculation of a kinetic energy for a known solution~\eqref{eq:el_anzats} is trivial, and ends up leading to the trivial expression for the mode energy
\begin{equation}
    \mathcal{H} = \rho \omega^2 \bm{u}^\ast\cdot\bm{u}.
    \label{eq:phon_norm}
\end{equation}

\section{Analytical treatment of elastic modes in a ring}
\label{elastic_ring_analytical}

Whereas the geometry we use in the main text, a ring on a stem, does not allow for an analytical solution, and has been solved numerically, the analytical solution is possible for a flat ring --- a thin disk of the radius $R$ with a middle hole of the radius $r_0$. The solutions for a disk without a hole can be produced by taking the limit $r_0 = 0$. The analytical solution for a ring will help us to understand the more complicated geometry of a ring with a stem.

The eigenfrequencies for the ring are given by a system of characteristic equations with the boundary conditions $\partial u/\partial \bm{n}=0$ at the boundaries. The deformation for $m=0$ produced by the linear combination of the Bessel functions, $u=J_1(\nu r)+C Y_1(\nu r)$. For disks it gives the inverse radius dependence $\omega \propto 1/R$ of the frequency, which also works well for $R\gg r_0$ in rings. The corresponding modes are shown as solid lines in Fig.~\ref{fig:mechanics-modes}. In this geometry, longitudinal and transverse modes are two distinct families of modes. The lowest frequency mode is the radial breathing mode.

\section{Vortex core contribution to the coupling}
\label{appendix:vortex_core}

For completeness, we provide here an estimate for the contribution of the vortex core to the coupling. Whereas the vortex core is not present in the ring geometry, which is the focus of this Article, it is present in the disc geometry.

In the core region, $\theta$ and $m_{\text{amp}}$ significantly depend on $r$, and the leading term of Eq.\eqref{magnetostriction_interaction} is
\begin{equation}
    \mathcal{H}_\text{int}=\frac{B_1-B_2}{3}m_{0z}m_{dz}\left(\frac{u_r}{r}+\frac{\partial u_r}{\partial r}\right).
\end{equation}
We now can calculate the overlap in the vortex core approximating both $m_\text{amp}$ and $u_r$ with Bessel functions $J_1, Y_1$ which were found to give an excellent fit for observed numerical modes. Further taking into account the quickly decaying profile of $m_z$, commonly modeled as $m_{0z}=\exp(-r^2/r_c^2)$, where $r_c\approx l_{ex}$ is a radius of the magnetic vortex core~\cite{Feldtkeller}, we arrive at the following estimate,
\begin{equation}
    g_{ms}^{(0)}\approx\sqrt{\frac{4\gamma_e}{\omega \rho M_s}}\frac{B_1-B_2}{24R^4}\frac{\sqrt{\pi}k_{el} k_m r_{c}^3}{\sqrt{n(k_{el})n(k_m)}},
\end{equation}
where $n(k)=\int_0^1 J_1^2(k r) r dr\approx 1$. Taking the constants from a fit of the numerical distributions we estimate $g/2\pi\sim 1$~kHz. This is a very low number compared to the contribution of the periphery of the vortex, considered in the main text, and can be safely disregarded.

\section{Field dependence of the coupling}
\label{app:field_dep}

When bias magnetic field is applied, the out-of-plane magnetization $m_{0z}$ rises linearly until it becomes close to the saturation field. Thus, one would expect that $\sin2\beta\approx 2B/(4\pi M_s)\sqrt{1-B^2/(4\pi M_s)^2}$, which is antisymmetric with the applied field, and decays more rapidly than it initially increases. The calculated behaviour is even more abrupt, because the in-plane angle $\alpha$ and the eccentricity $\epsilon$ also vary with the applied bias. Additionally, small local minima can be seen when magnon modes are tuned resonantly to each other, and part of magnon energy is diverted into hybridized non-coupling degrees of freedom. Unfortunately, it is hard to go to the saturation field completely with the methods used in Mumax3 (see Appendix \ref{ap:mumax}), because the applied wide-band excitation is enough to push the particle over into single axial domain state. This regime is however impractical for applications.

\begin{figure*}
    \subfloat{%
        \includegraphics[width=0.24\textwidth]{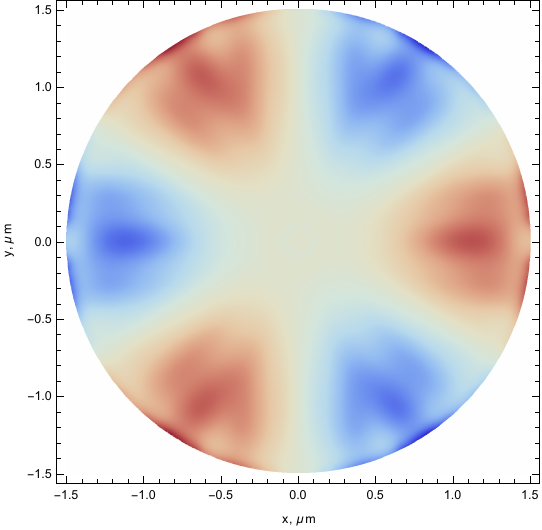}%
    }\hfill
    \subfloat{%
        \includegraphics[width=0.24\textwidth]{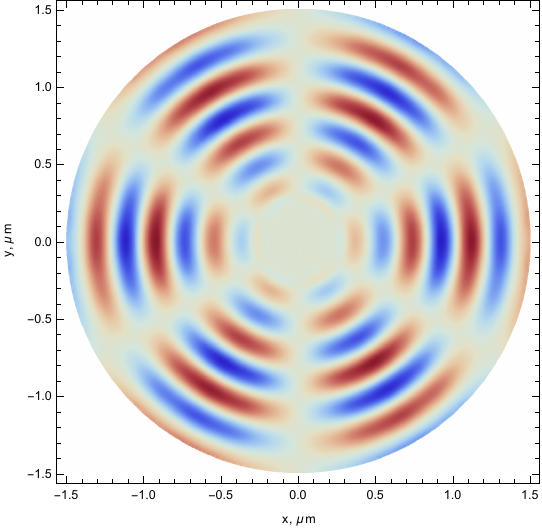}%
    }\hfill
    \subfloat{%
        \includegraphics[width=0.24\textwidth]{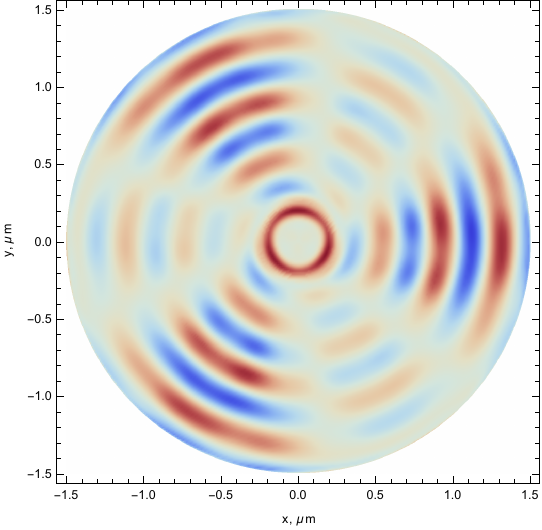}%
    }\hfill
    \subfloat{%
        \includegraphics[width=0.24\textwidth]{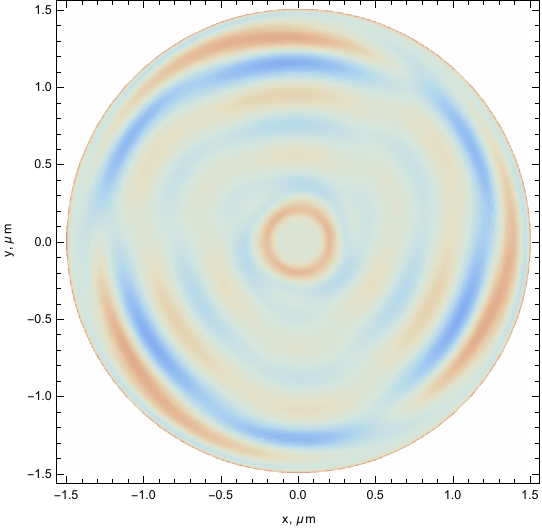}%
    }
    \\
    \centering\includegraphics[width=0.5\columnwidth]{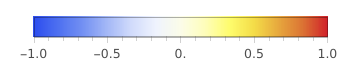}
    \caption{\label{fig:coupling-distribs}Instantaneous distribution of the energy density of the leading $B_2$ components of the magnetoelastic coupling~\eqref{magnetostriction_interaction}. The distribution (a) is that of a planar disk with no bias field. The other three distributions are for a realistic structure in an increasing bias field, no bias (b), weak bias (c), and strong bias (d) regimes are presented. The distributions for a real particle are planar projections of the coupling integral inside of the film, the density is taken from the height in the middle of a film locally. Comparing realistic nanostructures with disks, coupled transverse oscillations introduce additional nodes appearing along a radial direction. The external field is able to remove the inversion symmetry by deforming the static magnetization, and the opposite (in respect to the particle axis) azimuthal lobes no longer cancel out precisely. At large biases the magnetization is deformed to a degree where the overlap distribution changes qualitatively.}
\end{figure*}

How does the overlap cancel out is best demonstrated by the 3D numerical simulations done with Mumax3, Fig.~\ref{fig:coupling-distribs}. When the field is not applied, in the bulk of the particle the overlaps are significant, but after the integration over the area they cancel out almost completely (except for the core region as predicted by our previous calculations). In contrast, when the out-of-plane field is applied, the symmetry is removed, and the lobes on the opposite sides of the particle have different areas and different amplitudes, both contributing to uncompensated coupling.

\section{Mumax}
\label{ap:mumax}

All of the Mumax3 simulations mentioned were done with the custom geometry import function which has been pulled upstream with 3.11 release.

We reintroduce the cubic anisotropy with the constants $K_{c1} = \qty{-610}{\joule\per\meter^3}$ and $K_{c2}=\qty{-26}{\joule\per\meter^3}$~\cite{SUN2013157}. 

\begin{figure}[t]
\subfloat{\includegraphics[width=0.45\textwidth]{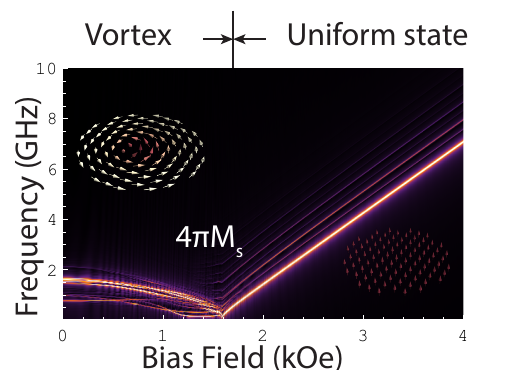}}
\caption{Simulated ferromagnetic resonance amplitude of the induced magnetic oscillations $\|m_z\|$ in a YIG ring ($t=\qty{100}{\nano\meter},\; R=\qty{1}{\micro\meter},\; r_0=\qty{200}{\nano\meter}$) excited by an in-plane external ac field as a function of the applied out-of-plane dc magnetic field $B$ done with Mumax3. The inset shows the magnetic vortex configuration. }
\label{fig:modes_mumax}
\end{figure}

In Mumax3 the damping was artificially increased to help with the convergence of simulations. We applied homogeneous out-of-plane dc bias magnetic field. The excitation is done with a broadband pulse of a $\sinc$ shape  $\sinc(\omega t)=\sin (\omega t)/(\omega t)$, $\omega$ being the high-frequency cutoff, to dramatically reduce the simulation time compared to a monochromatic excitation as all the modes are allowed to evolve at the same time and can be separated later using a Fourier transform. The amplitude of the ac pulse is taken to be spatially homogeneous as well. The ac pulse is linearly polarized along the x axis for the field dependence, Fig.~\ref{fig:modes_mumax}. The frequency of the pulse $\omega$ sets the upper limit on the frequency of the modes excited, and for our simulations value of $\omega=2\pi\times\qty{10}{\giga \hertz}$ was used, since all modes of interest are below this value. The amplitude of the pulse was selected such that the response is still linear.

As one can see from the Fig.~\ref{fig:modes_mumax} a good match is formed despite replacing the stem with a hole for Mumax3~\cite{Vansteenkiste}. This was a compromise we needed to take to reproduce the same amount of data to give smooth FMR spectra, since introducing a stem in Mumax3 is possible, but requires other compromises not compatible with the amount of data needed. However, Mumax3 serves as a good reference both for checking the eigenmodes, and the post-processing routines required to develop for the axial FEM solver to reproduce experimental observables.

When large datasets are needed, like the case for the FMR figure, we replace the complex geometry with disk with a hole which changes the picture by removing gyration modes since the vortex core area is cut out~\cite{PhysRevB.76.014416}, but doesn't much change the other, delocalized modes.

\end{document}